\documentclass[8.5pt,twoside,twocolumn]{article}
\usepackage[super,sort&compress,comma]{natbib} 
\usepackage{mhchem}
\usepackage{times,mathptmx}
\usepackage{sectsty}
\usepackage{balance} 

\usepackage{graphicx} 
\usepackage{lastpage}
\usepackage[format=plain,justification=raggedright,singlelinecheck=false,font=small,labelfont=bf,labelsep=space]{caption} 
\usepackage{fancyhdr}
\usepackage{color}
\usepackage{mathtools}

\graphicspath{{fig/}}

\begin{document}

\thispagestyle{plain}
\fancypagestyle{plain}{
\fancyhead[L]{\includegraphics[height=8pt]{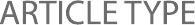}}
\fancyhead[C]{\hspace{-1cm}\includegraphics[height=20pt]{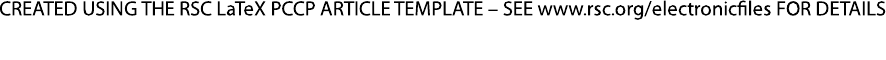}}
\fancyhead[R]{\includegraphics[height=10pt]{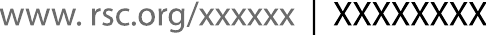}\vspace{-0.2cm}}
\renewcommand{\headrulewidth}{1pt}}
\renewcommand{\thefootnote}{\fnsymbol{footnote}}
\renewcommand\footnoterule{\vspace*{1pt}%
\hrule width 3.4in height 0.4pt \vspace*{5pt}} 
\setcounter{secnumdepth}{5}

\makeatletter 
\def\subsubsection{\@startsection{subsubsection}{3}{10pt}{-1.25ex plus -1ex minus -.1ex}{0ex plus 0ex}{\normalsize\bf}} 
\def\paragraph{\@startsection{paragraph}{4}{10pt}{-1.25ex plus -1ex minus -.1ex}{0ex plus 0ex}{\normalsize\textit}} 
\renewcommand\@biblabel[1]{#1}            
\renewcommand\@makefntext[1]%
{\noindent\makebox[0pt][r]{\@thefnmark\,}#1}
\makeatother 
\renewcommand{\figurename}{\small{Fig.}~}
\sectionfont{\large}
\subsectionfont{\normalsize} 

\fancyfoot{}
\fancyfoot[LO,RE]{\vspace{-7pt}\includegraphics[height=9pt]{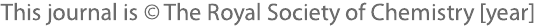}}
\fancyfoot[CO]{\vspace{-7.2pt}\hspace{12.2cm}\includegraphics{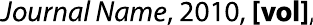}}
\fancyfoot[CE]{\vspace{-7.5pt}\hspace{-13.5cm}\includegraphics{headers/RF}}
\fancyfoot[RO]{\footnotesize{\sffamily{1--\pageref{LastPage} ~\textbar  \hspace{2pt}\thepage}}}
\fancyfoot[LE]{\footnotesize{\sffamily{\thepage~\textbar\hspace{3.45cm} 1--\pageref{LastPage}}}}
\fancyhead{}
\renewcommand{\headrulewidth}{1pt} 
\renewcommand{\footrulewidth}{1pt}
\setlength{\arrayrulewidth}{1pt}
\setlength{\columnsep}{6.5mm}
\setlength\bibsep{1pt}


\renewcommand{\vec}[1]{{\ensuremath{\mathchoice
                     {\mbox{\boldmath$\displaystyle\mathbf{#1}$}}
                     {\mbox{\boldmath$\textstyle\mathbf{#1}$}}
                     {\mbox{\boldmath$\scriptstyle\mathbf{#1}$}}
                     {\mbox{\boldmath$\scriptscriptstyle\mathbf{#1}$}}}}}%

\newcommand{\tens}[1]{{\ensuremath{\mathchoice
                     {\mbox{$\displaystyle\mathsf{#1}$}}
                     {\mbox{$\textstyle\mathsf{#1}$}}
                     {\mbox{$\scriptstyle\mathsf{#1}$}}
                     {\mbox{$\scriptscriptstyle\mathsf{#1}$}}}}}%

\newcommand{\csch}{\ensuremath{\mathrm{csch}}}

\newcommand{\Fvec}{\ensuremath{\vec{F}}}
\newcommand{\Svec}{\ensuremath{\vec{S}}}
\newcommand{\rvec}{\ensuremath{\vec{r}}}
\newcommand{\uvec}{\ensuremath{\vec{u}}}
\newcommand{\vvec}{\ensuremath{\vec{v}}}
\newcommand{\xvec}{\ensuremath{\vec{x}}}
\newcommand{\yvec}{\ensuremath{\vec{y}}}


\newcommand{\dcom}[2]{{\color{#1} #2}}
\newcommand{\dnew}[2]{{\color{#1} #2}}
\newcommand{\dcancel}[1]{{\sout{#1}}}
\newcommand{\dreplace}[3]{%
  {\dcancel{#1}{#2} $\Rightarrow$ \dnew{#1}{#3}}%
}

\newcommand{\US}[1]{{\dcom{blue}{#1}}}


\twocolumn[
  \begin{@twocolumnfalse}
\noindent\LARGE{\textbf{
    Collective waves in dense and confined microfluidic droplet arrays$^\dag$
}}
\vspace{0.6cm}

\noindent\large{\textbf{
    Ulf D. Schiller,\textit{$^{\ast}$$^{a}$}
    Jean-Baptiste Fleury,\textit{$^{b}$}
    Ralf Seemann,\textit{$^{bc}$} and
    Gerhard Gompper\textit{$^{a}$}  
}}\vspace{0.5cm}

\noindent\textit{\small{\textbf{Received Xth XXXXXXXXXX 20XX, Accepted Xth XXXXXXXXX 20XX\newline
First published on the web Xth XXXXXXXXXX 200X}}}

\noindent \textbf{\small{DOI: 10.1039/b000000x}}
\vspace{0.6cm}


\noindent \normalsize{%
  Excitation mechanisms for collective waves in confined dense one-dimensional microfluidic droplet arrays are investigated by experiments and computer simulations. We demonstrate that distinct modes can be excited by creating specific `defect' patterns in flowing droplet trains. Excited longitudinal modes exhibit a short-lived cascade of pairs of laterally displacing droplets. Transversely excited modes obey the dispersion relation of microfluidic phonons and induce a coupling between longitudinal and transverse modes, whose origin is the hydrodynamic interaction of the droplets with the confining walls. Moreover, we investigate the long-time behaviour of the oscillations and discuss possible mechanisms for the onset of instabilities. Our findings demonstrate that the collective dynamics of microfluidic droplet ensembles can be studied particularly well in dense and confined systems. Experimentally, the ability to control microfluidic droplets may allow to modulate the refractive index of optofluidic crystals which is a promising approach for the production of dynamically programmable metamaterials.}
\vspace{0.5cm}
  \end{@twocolumnfalse}
]


\footnotetext{\textit{
    $^{a}$~Theoretical Soft Matter and Biophysics, Institute of Complex Systems, Forschungszentrum J\"ulich, 52425 J\"ulich, Germany
}}
\footnotetext{\textit{
    $^{b}$~Experimental Physics, Saarland University, 66123 Saarbr\"ucken, Germany
}}
\footnotetext{\textit{
    $^{c}$~Max Planck Institute for Dynamics and Self-Organization, 37077 G\"ot\-tingen, Germany
}}


\footnotetext{\textit{
    $^{\ast}$~E-mail: u.schiller@ucl.ac.uk, Present address: Centre for Computational Science, University College London, 20 Gordon Street, London WC1H 0AJ, United Kingdom
}}

\footnotetext{
  \dag~Electronic Supplementary Information (ESI) available: Simulation movies showing collective longitudinal and transverse oscillations in arrays of drop\-lets. See DOI: 10.1039/b000000x/
}

\section{Introduction}

Microfluidic devices have become an important tool in chemistry and
biology, where they are increasingly used, for example, in analytic
essays,\cite{Huebner2008,Koester2008} micro-reactions \cite{Stone2004}
or flow cytometry.\cite{Mao2009} These applications typically involve
manipulation and control of immersed objects, such as droplets,
vesicles or cells,\cite{Squires2005} that interact hydrodynamically
through the flow perturbations of the surrounding fluid. A detailed
understanding of the correlated motion induced by long-ranged
hydrodynamic interactions in microfluidic devices is therefore
essential for efficient control of the flow of micro-particles. In
recent years, a number of studies brought out that microfluidic
droplet systems are especially well suited to steer their dynamics by
modifying particle properties and/or device
geometry.\cite{Baron2008,Baroud2010,Theberge2010,Seemann2012}
Consequently, microfluidic droplets have become both a test-bed and a
model system to study collective behaviour and self-organisation in
non-equilibrium many-body systems.\cite{Beatus2012} Typically, a
pressure-driven flow is imposed such that the system is out of
equilibrium, and at low Reynolds number viscous dissipation dominates
over inertia. A theoretical description of such driven dissipative
systems remains challenging,
and thus experiments and computer simulations are basic tools to study
the dynamics of microfluidic droplet ensembles.

When microfluidic droplets are confined between two parallel plates,
the geometry is effectively two-dimensional (2D) and the scattered
flow has a characteristic dipolar form.\cite{Diamant2009} In this
case, the hydrodynamic interactions are marginally long-ranged, i.e.,
the decay exponent is equal to the dimensionality of the
system.\cite{Campa2009} In contrast to quasi-1D geometries, where the
hydrodynamic interactions are strongly screened, the dipolar
interactions in quasi-2D geometries lead to complex collective
phenomena.\cite{Beatus2012} Dipolar flow fields are also
characteristic for some types of self-propelled particles, such as
droplets driven by Marangoni flows or by chemical reactions on their
surface.\cite{ChemCompBookChapter}
Some progress has been made in understanding the dynamics of rigid and
deformable particles and their hydrodynamic coupling in 2D
pressure-driven flow. Pairs of rigid particles in Poiseuille flow were
shown to follow either bound or unbound trajectories, depending on the
relative position of the particles, their absolute position in a
channel, and the strength of confinement.\cite{Reddig2013}
Linear arrays of rigid spheres and deformable drops aligned in the
flow direction undergo a pairing instability.\cite{Janssen2012} While
arrays of spherical particles are also unstable to lateral
perturbations, droplet arrays are stabilised by quadrupolar
interactions due to deformation.\cite{Janssen2012}
Asymmetric particles align with the flow due to self-interactions, and
migrate to the centreline of the confining
channel.\cite{Uspal2013,Adler1981,Ekiel-Jezewska2009} For highly
asymmetric particles, the time-scales for alignment and focusing
separate due to the distinct hydrodynamic mechanisms involved. The
focusing of asymmetric particles resembles a damped harmonic
oscillator, whereas symmetric particles oscillate between
side-walls.\cite{Uspal2013}
In 2D ensembles of droplets, the dipolar hydrodynamic interactions
give rise to sound and shock waves that are superposed on droplet
diffusion. The waves are due to a density-velocity coupling and can be
described by a 1D Burgers equation.\cite{Beatus2009,Beatus2012}

A particular feature arises in regular arrays of droplets, so called
microfluidic crystals, where the flowing droplets have a spatial order
with a well-defined spacing. These crystals can exhibit collective
oscillations with a dispersion relation akin to solid state
phonons.\cite{Beatus2006,Beatus2007} These microfluidic phonon modes
are neither growing nor decaying, and are thus a realisation of
marginally stable oscillatory modes in a dissipative system made
possible by the imposed symmetry-breaking flow. Practically, however,
the possibility to observe these modes is limited by non-linear
instabilities \cite{Beatus2012,Liu2012} and the strong dependence on
initial conditions. Only recently, an experimental technique was
proposed to systematically excite microfluidic phonons, and the
observed modes revealed a coupling mechanism, induced by lateral
confinement, between longitudinal and transverse modes that was
confirmed by computer simulations.\cite{Fleury2014}
The ability to control the dynamic properties by tuning the flow
characteristics opens interesting perspectives regarding dynamically
programmable metamaterials which could be produced by modulating the
refractive index of droplet
crystals.\cite{Hashimoto2006,Chen2012,Uspal2014}

Here we investigate collective modes in dense microfluidic crystals
under confinement both experimentally and by computer simulations. We
show that distinct oscillatory behaviour can be systematically excited
by varying the initial conditions through the introduction of specific
`defect' patterns. The observed modes are ana\-lysed and
characterised, and reveal several interesting dynamic features, such
as cascades of laterally offset pairs and mode coupling. The results
from experiments and computer simulations agree quantitatively. The
long-time behaviour is investigated in computer simulations and used
to identify possible instabilities and their underlying
mechanisms. Our approach demonstrates the rich dynamics that emerges
from hydrodynamic interactions in confined microfluidic droplet
ensembles. The results show good agreement with a linearised far-field
theory \cite{Beatus2012} even in the dense droplet regime. This makes
it very promising to apply the techniques to other crowded
microfluidic systems, such as self-propelled
particles.\cite{ChemCompBookChapter}

The remainder of the article is organised as follows: In section
\ref{sec:theory}, we review the hydrodynamics of quasi-2D systems and
the linearised far-field theory for microfluidic phonons. Section
\ref{sec:methods} describes the experimental techniques and the
simulation approach we used to study microfluidic droplet systems. In
section \ref{sec:excitation}, we present excitation mechanisms for
collective waves and analyse the observed oscillations and
instabilities. A concluding discussion is given in section
\ref{sec:conclusions}.


\section{Microdroplet trains in quasi-2D flow}
\label{sec:theory}

\begin{figure}
\centering
\includegraphics[width=\columnwidth]{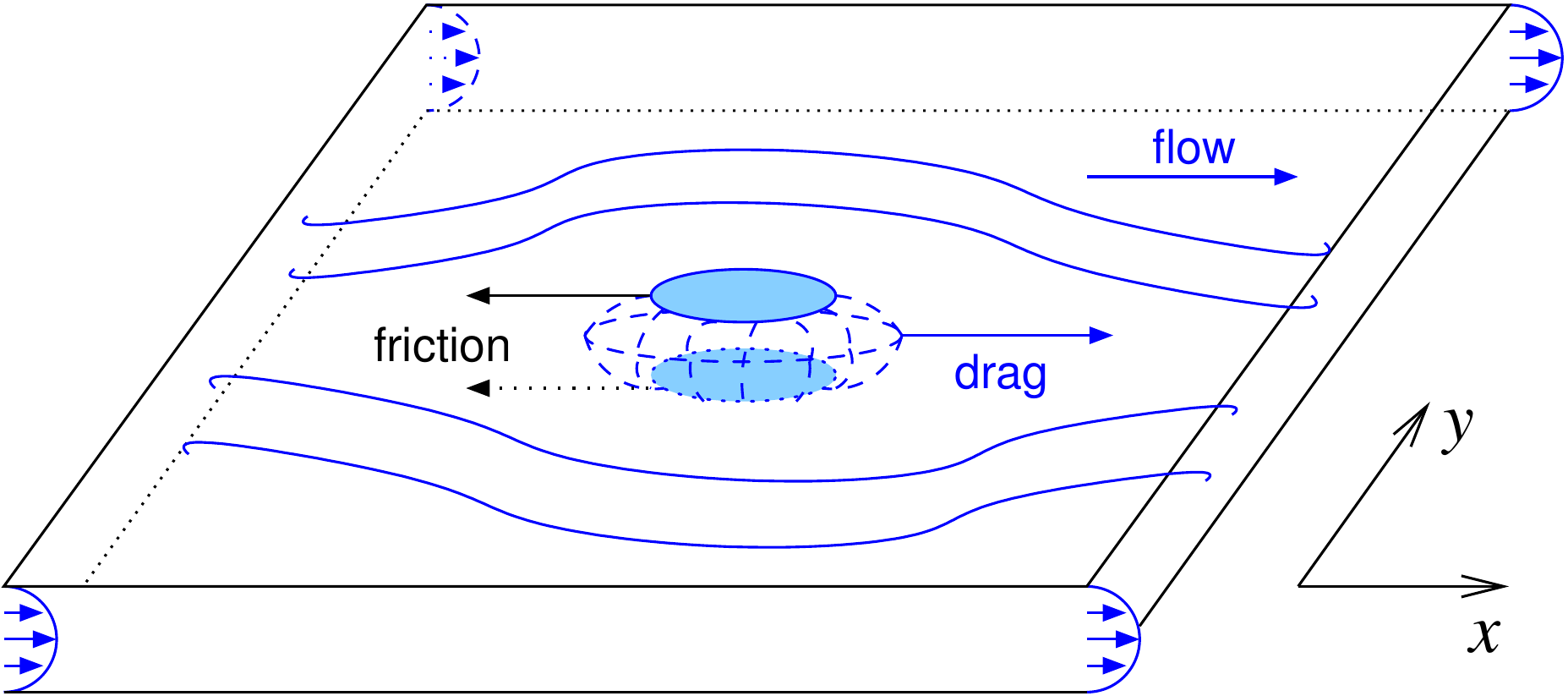}
\caption{Schematic illustration of the quasi-2D flow geometry. Flattened
  droplets experience a friction at the top and bottom plates which
  counteracts the hydrodynamic drag. When the droplets move relative
  to the imposed flow, they act as a mass dipole with a sink at their
  leading edge and a source at their trailing edge.}
\label{fig:hele-shaw}
\end{figure}

We consider droplets that are confined between two parallel plates and
thus move in a quasi-2D geometry, cf. figure \ref{fig:hele-shaw}. The
flow and the hydrodynamic interactions in this geometry differ
qualitatively from the bulk case due to momentum absorption at the
confining plates which leads to screening of the far-field. The fluid
flow satisfies no-slip boundary conditions on the channel walls, and
since the height $H$ of the channel is small compared to the lateral
width $W$, the velocity gradient in the $z$-direction is much larger
than in the planar directions. In the Darcy approximation
$\partial_z^2 \gg \partial_x^2$, $\partial_z^2 \gg \partial_y^2$, the
solution of the Stokes equation has a quasi-2D Hele-Shaw form
\cite{Bensimon1986}
\begin{equation}\label{eq:hele-shaw}
\begin{split}
\vec{u}(x,y) 
&= \frac{1}{H} \int_{-H/2}^{H/2} dh \frac{h^2-H^2/4}{2\eta} \nabla p(x,y)  \\
&= -\frac{H^2}{12\eta} \nabla p(x,y) .
\end{split}
\end{equation}
At low Reynolds number, the flow is incompressible and
Eq.~(\ref{eq:hele-shaw}) can be written as a Laplace equation
\begin{align}
\nabla^2 \phi(x,y) &= 0 \\
\uvec(x,y) &= \nabla \phi(x,y)
\end{align}
where the effective potential is defined through the pressure $\phi =
-H^2p/12\eta$ with $\eta$ the dynamic viscosity of the fluid.

We briefly review here the theoretical description presented in
Refs.\cite{Beatus2007,Beatus2012} When a droplet is moving through the
fluid with a velocity $\delta \vec{u} = \vec{u}^\infty - \vec{u}_d$
relative to the externally imposed flow $\vec{u}^\infty = u^\infty
\hat\xvec$, it acts as a momentum monopole whose flux scales as
$\delta u^2$. However, due to the absorption of momentum at the top
and bottom plates the flux is not conserved. The absorbed flux scales
as $\delta u/h$, therefore the flow field of the momentum monopole
$\partial_r \delta u \propto - \delta u/h$ decays
exponentially.\cite{Diamant2009} Thus, unlike in the bulk case, the
leading contribution is the mass dipole created by the droplet. The
moving droplet pushes fluid out at the upstream edge and draws fluid
in at the downstream edge, giving rise to a characteristic dipolar
flow field.
Formally, the flow perturbation at a distance $\vec{r}$ from a droplet
is obtained by solving the Laplace equation with boundary conditions
of zero mass flux (zero perpendicular velocity) through the droplet
interface.\cite{Tlusty2006} This gives the dipolar potential and
scattered velocity field around a droplet of radius $R$
\begin{align}\label{eq:dipole1}
\phi_d(\rvec) &= R^2 \delta \vec{u} \cdot \frac{\hat\rvec}{r} \\
\uvec(\rvec)  &= R^2 \delta \vec{u} \cdot \frac{\mathsf{I} - 2 \hat\rvec \hat\rvec }{r^2} .
\end{align}
%
%
The 2D potential flow can also be described by a complex potential
$w(x+iy)=\phi(x,y)+i\psi(x,y)$ where the imaginary part is the stream
function $\psi(x,y)$. The flow velocity is then given by $u_x - i u_y
= dw/dz$ with $z=x+iy$. For an imposed flow in the $x$-direction the
complex dipolar potential is then
\begin{equation}\label{eq:dipole}
w_d(z) = R^2 \delta u \frac{1}{z} .
\end{equation}

A droplet moving in the imposed flow experiences a hydrodynamic
drag that can be written as 
\begin{equation}
F_\text{h} 
= \frac{1}{2} \xi u_d + \frac{\xi}{R^2} \sum \mathrm{Res}[w] 
= \frac{1}{2} \xi u_d + \xi \delta u ,
\end{equation}
where the drag coefficient $\xi=24 \pi \eta R^2/H$ is introduced.
The second term arises from the self-interaction of the droplet with
its dipole.\cite{Beatus2012}

If the size $R$ of the droplets exceeds the channel height $H$, they
are flattened and experience a friction with the top and bottom plates
which can be modelled as
\begin{equation}\label{eq:friction}
\vec{F}_\text{f} = - \zeta \vec{u}_d .
\end{equation}
Since inertial effects can be neglected at low Reynolds number, we can
use force balance $F_\text{h}+F_\text{f} = 0$ to obtain the equation of
motion for the droplet
\begin{equation}
\vec{u}_d = \mu \vec{u}^\infty 
= \left( \frac{1}{2} + \frac{\zeta}{\xi} \right)^{-1} \vec{u}^\infty ,
\end{equation}
where $\mu = u_d / u^\infty$ is the mobility of the droplet in the
imposed flow.

In the presence of lateral side-walls, i.e., in a microfluidic
channel, additional boundary conditions have to be satisfied. The
simple dipole potential~(\ref{eq:dipole}) has a non-vanishing flux at
the side-walls which can be eliminated by placing image dipoles inside
the wall.\cite{Lamb1932,Milne-Thomson1962} These dipoles form an
infinite array perpendicular to the flow direction. The flow potential
of a single droplet is obtained by summing over the infinite dipole
array, and then rescaling by a compressibility factor $C$ to account
for the finite size $R$ of the droplets.\cite{Lamb1932,Beatus2012} The
result is \cite{Beatus2007}
\begin{equation}\label{eq:warray}
\begin{multlined}
w_d(z) =
C \cdot \frac{\pi R^2 \delta u}{2W} 
\left\{
  \coth \left[ \frac{\pi}{2W} \left( z - i y_d \right) \right]\right.\\
\left.+ \coth \left[ \frac{\pi}{2W} \left( z - i (W - y_d) \right) \right]
\right\} ,
\end{multlined}
\end{equation}
where
\begin{equation}
C = \frac{2W}{\pi R} \left[
\cot \left( \frac{\pi R}{2W} \right)
- \frac{\sin\left(\frac{\pi R}{W}\right)}
  {\cos\left(\frac{\pi R}{W}\right) + \cos\left(\frac{2\pi y_d}{W}\right)}
\right]^{-1} .
\end{equation}
The equation of motion for a confined droplet can be obtained as above
and keeps the form $u_d = \mu u^\infty$ if the mobility is replaced by
\begin{equation}
\mu = C \cdot \left( C - \frac{1}{2} + \frac{\zeta}{\xi} \right)^{-1} .
\end{equation}

In an ensemble of droplets, the solution of the Laplace equation is
considerably more complicated because the boundary conditions have to
be satisfied additionally on all droplet surfaces. Although this is in
principle possible using the method of images, the large number of
reflections that arise makes it unreasonably intricate in
practice. One therefore resorts to the leading-order approximation
where the drag force is given by a superposition of the flow fields
created by the other droplets.
For the $n$-th droplet in an ensemble, the equation of motion thus is
\cite{Beatus2012}
\begin{equation}\label{eq:ensemble}
u_{n,x} - i u_{n,y} = \mu \left( u^\infty + \sum_{j \neq i} 
\left.\frac{d w_d}{d z} \right|_{z_j-z_n} \right) .
\end{equation}
This approximation is valid if the inter-droplet distance is larger
than the droplet size $r_j-r_n \gg R$, and we will see below that the
predictions based on Eq.~(\ref{eq:ensemble}) work well even for dense
droplet trains.

For a regular train of droplets flowing with an `equilibrium'
spacing $a$ in the centre of the channel, the displacements $\delta
z_n = z_n - n a$ are assumed to be small and the derivative of the
potential (\ref{eq:warray}) can be expanded. To first order, the
equations of motion are then given by \cite{Beatus2012,Fleury2014}
\begin{equation}
\begin{split}
\dot{\delta x_n} &=
\begin{multlined}[t]
2 B \sum_{j=1}^{\infty} \coth \left(\frac{a j \pi}{W}\right)\\
\times \csch^2 \left(\frac{a j \pi }{W}\right) \left( \delta x_{n+j} - \delta x_{n-j}
\right)
\end{multlined}\\
\dot{\delta y_n} &= 
\begin{multlined}[t]
- B \sum_{j=1}^{\infty} \left[ 1 + \cosh\left(\frac{a j \pi}{W}\right) \right]^2 \\
\times \csch^3 \left(\frac{a j \pi }{W}\right) \left( \delta y_{n+j} -
\delta y_{n-j} \right) ,
\end{multlined}
\end{split}
\end{equation}
with prefactor $B = \mu (u^\infty-u_d) (\pi^2R/W^2)\tan(\pi R/W)$.
These equations describe waves, and by plugging in a plane-wave
solution, we arrive at the dispersion relations
\begin{equation}\label{eq:dispersion}
\begin{split}
\omega_{\parallel}(k) &= -4 B \sum_{j=1}^{\infty}
\coth\left(\frac{a j \pi}{W}\right) 
\mathrm{csch}^2\left(\frac{a j \pi}{W}\right) \sin(j k a) \\
\omega_{\perp}(k) &= 
\begin{multlined}[t]
2 B \sum_{j=1}^{\infty}
\left[ 1 + \cosh\left(\frac{a j \pi}{W}\right) \right]^2 \\
\times \mathrm{csch}^3\left(\frac{a j \pi}{W}\right) \sin(j k a) .
\end{multlined}
\end{split}
\end{equation}
%


\section{Experimental and computational droplet microfluidics}
\label{sec:methods}

Having discussed the flow fields and the forces acting on a single
droplet, we discuss the experimental realisation as well as computer
simulations of flowing droplets in a straight microfluidic channel.

\subsection{Microchip fabrication and droplet production}
\label{sec:exp}

Microfluidic devices were fabricated in Sylgard 184 (Dow Corning)
using standard soft litho\-graphic
protocols,\cite{Duffy1998,Chokkalingam2008} and the flow rates were
volume-controlled by syringe pumps. Mono-disperse water droplets were
generated in n-hexadecane ($\rho=773\,\text{kg/m}^3$, $\eta =
3\,\text{mPa\,s}$) with 2\,wt\% of the surfactant Span\,80 using a
step geometry,\cite{Priest2006,Chokkalingam2008}. The microchannel has
uniform height and width of $H \times W \approx 120\,\mu\text{m} \times
210\,\mu\text{m}$.
%
Typical flow velocities are $u_{d} \approx 250\,\mu\text{m/s}$ for the
droplet, and $u_{oil} \approx 500\,\mu\text{m/s}$ for the continuous
oil phase.  The corresponding Reynolds and Peclet number are $Re =
\rho u_\text{oil} R/\eta \approx 10^{-2}$ and $Pe = u_\text{oil} R/D
\approx 10^{8}$, respectively. 

\subsection{Simulation approach: Multi-particle collision dynamics}
\label{sec:mpc}

In order to investigate the origin of our experimental observations
and the validity of the approximations in the theoretical description,
we conduct computer simulations using multi-particle collision
dynamics (MPC). MPC is a mesoscopic simulation method that is capable
of reproducing the full hydrodynamics of a
fluid.\cite{Malevanets1999,Kapral2008,Gompper2009} Since it does not
rely on the assumption of a specific flow perturbation, it is well
suited to test the accuracy of the semi-analytical theory based on dipolar
flow fields to describe the droplet interactions in a dense and
confined system.
The fluid is modelled explicitly by idealised point-like particles of
mass $m$. The fluid dynamics emerges from local mass, momentum and
energy conservation in the particle ensemble, whose equation of state
is that of an ideal gas. The update of particle positions and momenta
mimics the underlying kinetics and is split into successive streaming
and collision steps.  During the streaming step the particle moves
ballistically,
\begin{equation}
\rvec_i = \rvec_i + h \vvec_i ,
\end{equation}
where $h$ is the time interval between collisions. In the collision
step, the particles are sorted into cubic collision cells of size
$\Delta x$. In each cell, the particles then exchange momentum while
the momentum of the collision cell is conserved. Various collision
rules have been proposed in the literature and in this work, we employ
a collision rule that also conserves angular momentum of the
cell.\cite{Goetze2007} The collisions are augmented with an
Anderson-like thermostat to control the temperature. The overall
update of particle velocities is
\begin{equation}
\begin{multlined}
\vvec_i^* =
\vvec_C + \vvec_i^\text{ran} - \sum_{j\in C} \frac{\vvec_j^\text{ran}}{N_C} \\
+ m \tens{\Pi}^{-1} \sum_{j\in C} \left[ \rvec_{j,C} \times \left( \vvec_j - \vvec_j^\text{ran} \right) \right]  \times \rvec_{i,C} ,
\end{multlined}
\end{equation}
where $\vvec_C$ is the centre of mass velocity of the collision cell
containing $N_C$ particles, $\tens{\Pi}$ is the moment of inertia
tensor of the particles, $\rvec_{i,C}=\rvec_i - \rvec_C$ is the
relative particle position, and $\vvec_i^\text{ran}$ is a random
velocity drawn from a Maxwell-Boltzmann distribution. This collision
operator is denoted as MPC-AT+$a$ in the nomenclature of
Ref.~\cite{Noguchi2008}. In addition, the cell grid is shifted
randomly before each collision step to restore Galilean invariance of
the system.\cite{Ihle2001} The dynamic viscosity $\eta$ of the
MPC-AT+$a$ fluid for large number density $n$ (particles per cell) is
then given by
\begin{equation}
\eta = \frac{n k_BT h}{\Delta x^d} \left( \frac{n}{n-(d+2)/4}-\frac{1}{2} \right)
+ \frac{m (n-7/5)}{24 \Delta x^{d-2} h} ,
\end{equation}
where $k_BT$ is the imposed temperature and $d=2$ is the
dimensionality of the system.

Since the droplets hardly deform in the experiment, we model them as
rigid discs of radius $R$ that are coupled to the fluid by a no-slip
boundary condition, i.e., $\vvec' = -\vvec + 2\vvec_\text{b}$ where
$\vvec_\text{b}$ is the boundary velocity. It is to be noted that this
is effectively a different boundary condition than the one used in
deriving Eq.~(\ref{eq:dipole1}), however, we have found in practice
that this can be accounted for by the calibration procedure described
below and does not lead to a relevant difference in the
measurements. To apply the collision rule in the cells that are partly
or fully occupied by the rigid discs, the corresponding volume is
filled with virtual particles that are distributed randomly within a
layer of width $\sqrt{2}a$ and whose velocities are distributed
according to a Maxwell-Boltzmann distribution around the boundary
velocity $\vvec_\text{b}$.\cite{Goetze2007} The momentum change of the
fluid particles during streaming and collisions is accumulated and
leads to the boundary force $\vec{F}_\text{b}$ that moves the
discs.\cite{Goetze2011}

On very short distances, hydrodynamics are not resolved and hence we
add steric droplet-droplet and droplet-wall interactions by means of
Weeks-Chandler-Anderson (WCA) potentials
\begin{equation}
V_\text{WCA}(r) =
4\epsilon \left[ \left( \frac{r_0}{r} \right)^{12} - \left(
\frac{r_0}{r} \right)^6 \right] + \epsilon \qquad 0 < r < 2^{\frac{1}{6}} r_0 ,
\end{equation}
where $r = r_{ij} - 2R$ for droplet-droplet interactions, and $r =
r_{ij} - R$ for droplet-wall interactions. $r_{ij}$ is the distance of
the droplet centres, or the distance of the droplet centre from the
wall surface, respectively. To account for the friction at the top and
bottom of the Hele-Shaw cell, cf. Eq.~(\ref{eq:friction}), we apply a
friction force $\vec{F}_\text{friction} = -\gamma \, \vec{u}_\text{d}$ to
the discs where $\vec{u}_\text{d}$ is the velocity of the droplet
relative to the microchannel. The overall flow is driven by an
external force $\vec{g}$ corresponding to a constant pressure gradient
across the channel.

The parameters of the MPC simulations are as follows. The size of a
collision cell is $\Delta x=R/5$ and the time step is $h =
0.005\,\tau$, where the time scale is $\tau=(k_BT/m)^{-1/2} \Delta x$,
and $m$ is the mass of one MPC particle. The fluid density, the
driving force and the friction are $\rho=40\,m/\Delta x^2$, $g=0.1\,m
\Delta x/\tau$, and $\gamma=2 \cdot 10^4\,m/\tau$. These parameters
correspond to a fluid viscosity $\eta=321.77\,m/\tau$. The mass of the
droplets is given by $M=\rho\pi R^2\approx3140\,m$. The parameters for
the WCA potential are $\epsilon=k_BT$ and $r_0=\Delta x$. We varied
the channel width $W$, the droplet spacing $a$, and the initial
configuration including the initial wavelength $\lambda$ for
simulating droplet trains in a channel.

In order to compare the simulation results to the experiments
quantitatively, we determined the value $K = u_d/u_\text{oil}$ from
independent simulation runs with a single droplet under the same
confinement. For a channel width of $W=3R$ we obtained a value of $K
\approx 0.62$ which is on the order of the experimental parameters.
The oil velocity is in the range $u_\text{oil} \approx 10^{-2} \Delta
x/\tau$, such that the typical Reynolds and Peclet number of the
simulations are $Re \approx 10^{-3}$ and $Pe \approx 10^2$,
respectively. Note that the Peclet number is significantly lower than
in the experiments. On the one hand, this leads to more pronounced
fluctuations, but on the other hand, it allows us to observe in the
simulations the onset of instabilities on accessible time scales,
cf. section \ref{sec:stability}.

%
%


\section{Controlled excitation and analysis of collective oscillations}
\label{sec:excitation}

Arrays of droplets flowing in a straight microchannel self-organise
into two parallel trains of droplets with alternating lateral
positions.\cite{Thorsen2001} For all neighbours $j$ of a droplet $i$
in this zigzag order, there exists a droplet $j'$ such that the
positions relative to $i$ satisfy $\rvec_{ij} \cdot \hat\xvec = -
\rvec_{ij'} \cdot \hat\xvec$ and $\rvec_{ij} \cdot \hat\yvec =
\rvec_{ij'} \cdot \hat\yvec$. Therefore, the flow fields of droplets
$j$ and $j'$ cancel at the position of droplet $i$ and no force is
exerted due to the symmetry of the arrangement. The zigzag order is
thus stable,
%
%
%
and for a collective oscillation to emerge the symmetry of the droplet
arrangement has to be broken. 
In the following, we describe ways to excite oscillations in an array
of droplets and analyse quantitatively the collective modes observed
in experiments and reproduced by computer simulations using MPC as
described in section \ref{sec:mpc}.
%
%
%
%
The results demonstrate the rich dynamics that emerge from
hydrodynamic interactions in confined microfluidic droplet ensembles.


\subsection{Longitudinal oscillations}
\label{sec:long}

\begin{figure}
\includegraphics[width=\columnwidth]{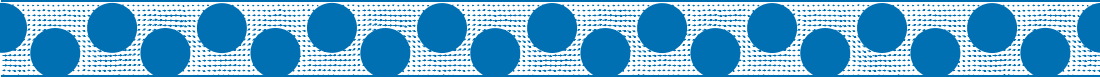}\smallskip\\
\includegraphics[width=\columnwidth]{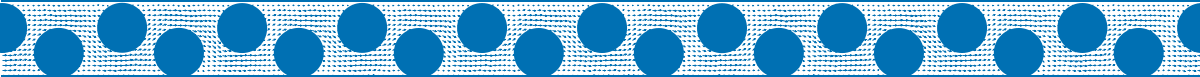}\smallskip\\
\includegraphics[width=\columnwidth]{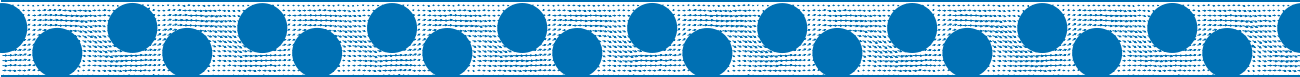}
\caption{Illustration of the introduction of gaps in the zigzag
  arrangement in computer simulations. The gaps lead to longitudinal
  pairing cascades along the train of droplets.$^\dag$ The average droplet
  spacing $a$ in the configurations shown from top to bottom is
  $2.2\,R$, $2.4\,R$, and $2.6\,R$.  }
\label{fig:gap_setup}
\end{figure}

\begin{figure}
\includegraphics[width=\columnwidth]{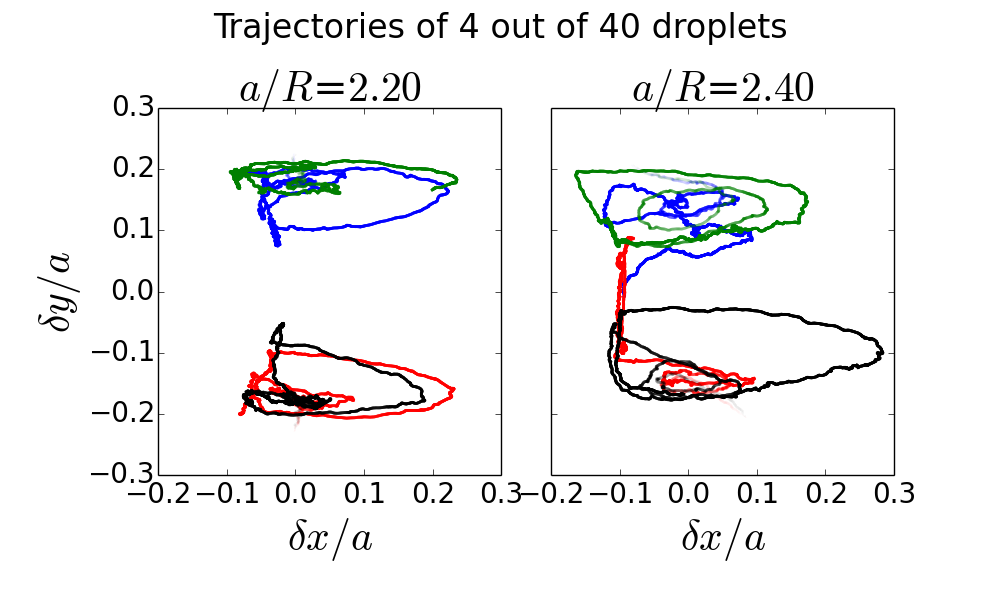}
\caption{Trajectories of $4$ droplets in a train of $40$ droplets
  obtained from simulations where the initial arrangement contains
  gaps. For the smaller spacing $a=2.2\,R$ the longitudinal and
  transverse oscillations remain stable for some time until the
  longitudinal oscillations start growing. The transverse oscillations
  do not cross the channel centreline, and the trajectory in
  configuration space has a circular form. For the larger spacing
  $a=2.4\,R$ the oscillations start to grow sooner and the droplets move
  transversely across the channel, thus breaking the initial pattern.}
\label{fig:trajectories-long}
\end{figure}

In simulations, one way to perturb the symmetry of a droplet train is
to vary the spacing by introducing `gaps' between the incoming
droplets in regular intervals.
The resulting arrangement for the case of an alternating sequence 
of two different distances between subsequent droplets
is illustrated in figure~\ref{fig:gap_setup}. The droplet crystal now
consists of pairs of laterally displaced droplets.
For an isolated pair of spherical droplets, the hydrodynamic forces do
not lead to relative particle motion,\cite{Janssen2012} owing to the
time reversibility in the Stokes regime. In an ensemble, however, the
pairs are also affected by the neighbouring pairs. In the perturbed
zigzag arrangement, the pairs are separated by a larger distance than
the droplets in the pair, and due to these gaps the configuration of the
ensemble to the right of any droplet is different from the
configuration on its left. Consequently, the flowing droplets can
experience a net hydrodynamic force and undergo relative motion.
\begin{figure}
\includegraphics[width=\columnwidth]{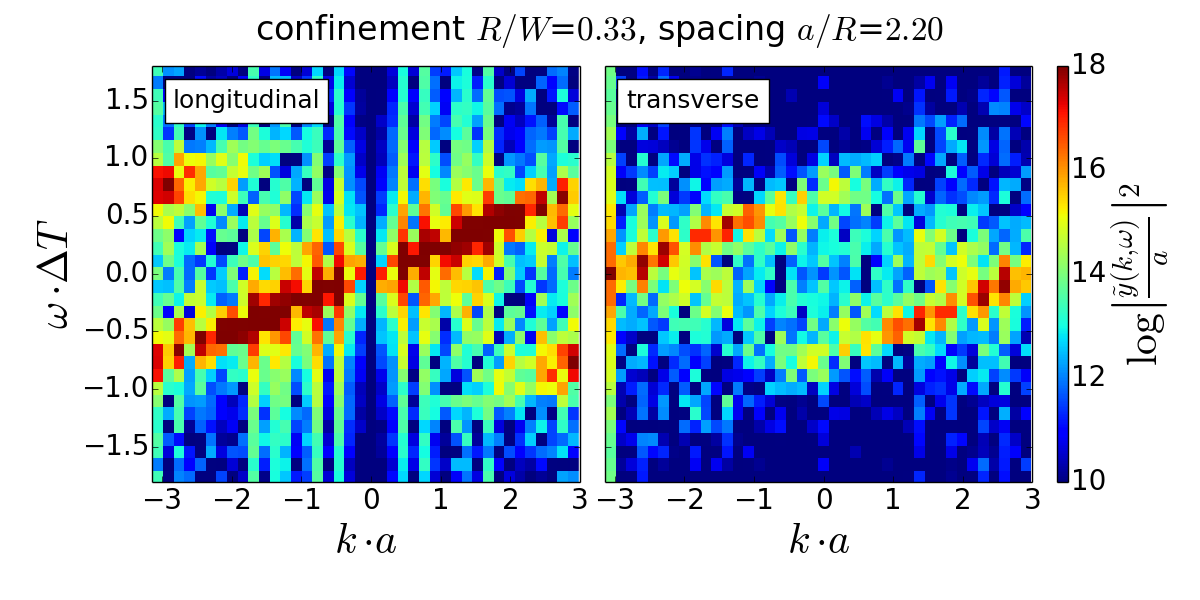}\\
\includegraphics[width=\columnwidth]{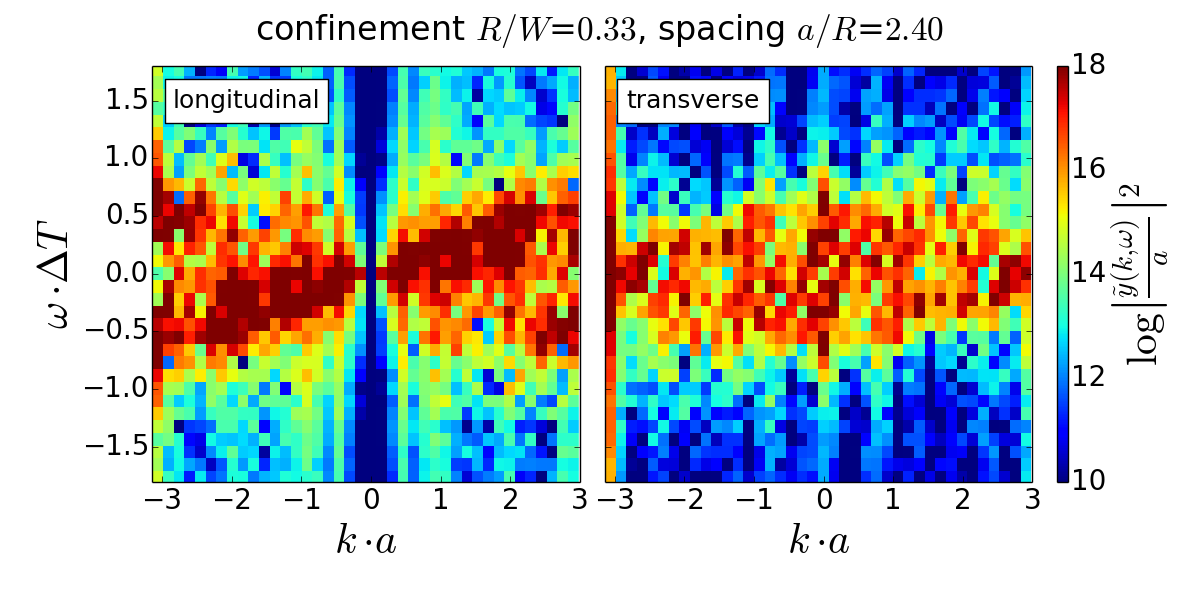}
\caption{Power spectra obtained from simulations of the oscillations
  triggered by the introduction of gaps.
}
\label{fig:spectra-long}
\end{figure}
We observe that the leading droplet of the pair moves faster than the
trailing droplet. It separates and catches up with the trailing
droplet of the pair ahead of the original one, thus forming a new
pair. This process repeats with the new pairs and creates a
longitudinal oscillation of the droplet distances in the train, where
adjacent droplets are in anti-phase.$^\dag$
Figure~\ref{fig:trajectories-long} shows the motion pattern for this
droplet arrangement in the co-moving frame of reference.
The behaviour shows some similarity to the pairing cascades observed
in finite droplet trains \cite{Janssen2012}, however, in an infinite
crystal the pairs cannot separate and keep interacting such that the
oscillatory motion ensues.  This is also reminiscent of the behaviour
of colloidal particles driven by a constant external pulling 
force to move on a ring.\cite{Lutz2006}
The motion patterns further reveal that in the initial stage, the
transverse amplitude is small and the droplets stay on one or the
other side of the channel. In this phase, the configuration space
trajectory of the droplets has a circular pattern. Over time, however,
the transverse amplitude is growing and the droplets eventually cross
the centre-line of the channel, as can be seen in figure
\ref{fig:trajectories-long} for $a/R=2.4$.

\begin{figure}
\includegraphics[width=\columnwidth]{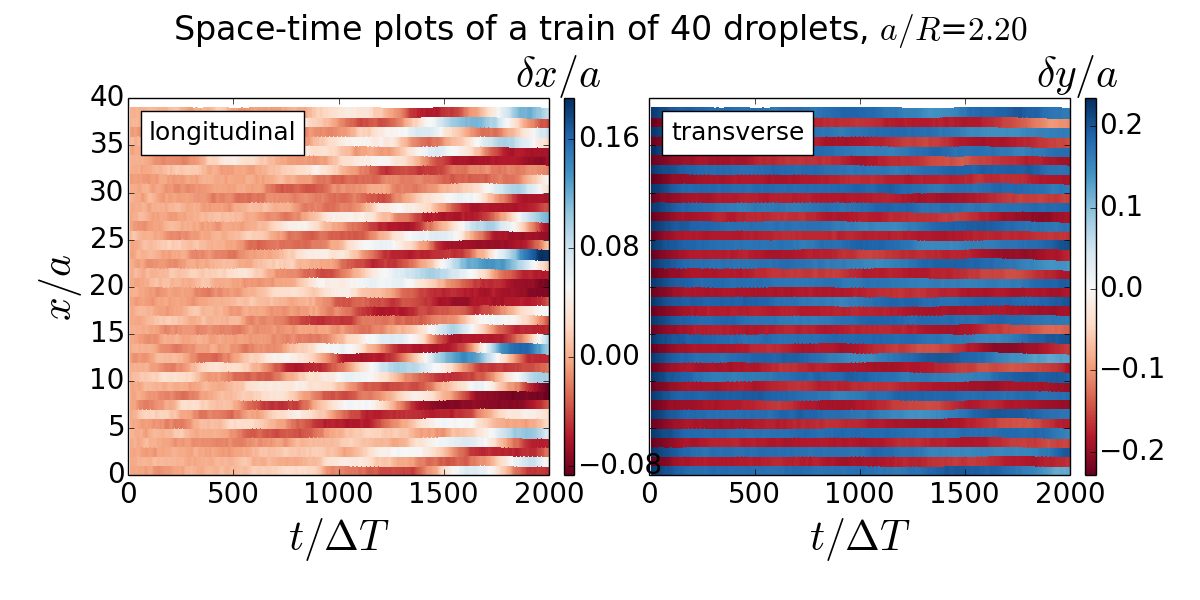}\\
\includegraphics[width=\columnwidth]{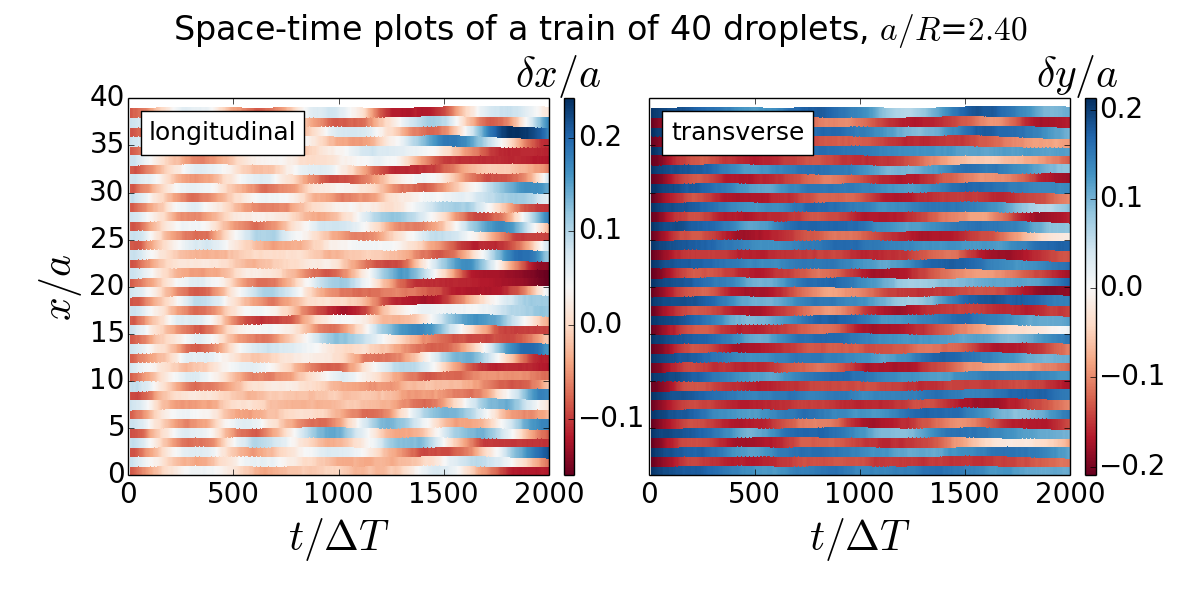}
\caption{Space-time plots of simulation runs of a train of $40$
  droplets perturbed by gaps. The initial horizontal pattern indicates
  that the oscillations start out as a standing wave. After some time
  the initial pattern breaks, and waves begin to propagate along the
  droplet train.}
\label{fig:spacetime-long}
\end{figure}

The longitudinal and transverse power spectra of the droplet
oscillations observed in simulations are shown in figure
\ref{fig:spectra-long}. For the smaller spacing $a=2.2\,R$, a clear
signature of longitudinal oscillations is observed. The accompanying
transverse oscillations are from droplets moving towards the centre of
the channel and back without crossing the channel. For larger spacing
$a=2.4\,R$ the power spectra show more scattered features without a
clear signature indicating the limited stability the gap modes.
The longitudinal spectrum has a strong signature $\omega \propto k$
which is due to fluctuations, but there are also distinct signals at
$k \cdot a = \pi$ indicating a zigzag wave. 
%

The space-time plot of the droplet distance in figure
\ref{fig:spacetime-long} shows for small times a pattern of horizontal
stripes in the both the longitudinal $\delta x$ and transverse $\delta
y$ displacements, which indicates that in the co-moving frame, the
excited mode is initially a standing wave. For longer times, the
pattern changes as a travelling wave seems to develop.
%
%
These results suggest that the stability of the longitudinal pairing
wave is limited, as we discuss in more detail below. 
%


\subsection{Transverse oscillations}
\label{sec:trans}

%
%
%
%
%
%

\begin{figure}
\includegraphics[width=\columnwidth]{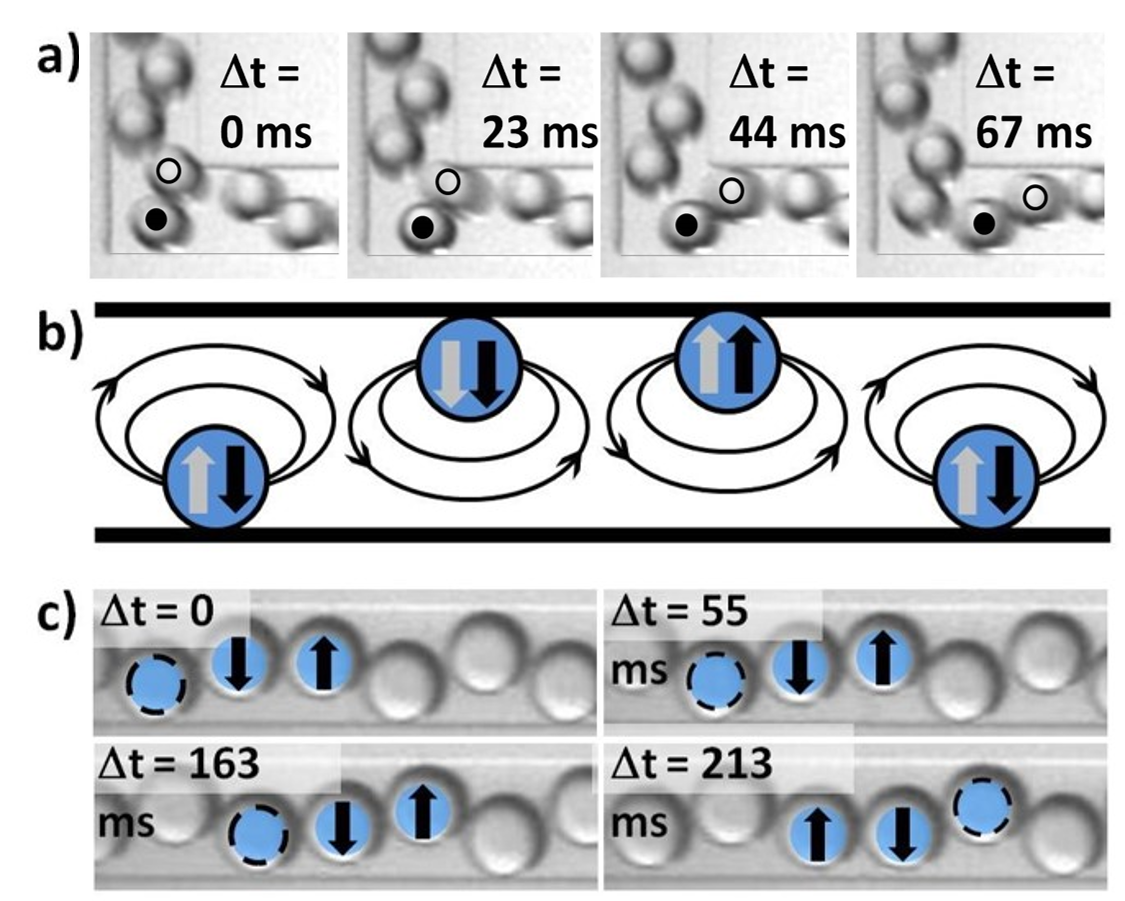}
\caption{
  a) Microscopy time series showing the droplet reorganisation at a
  $90^\circ$ bend.  b) Sketch indicating the relevant hydrodynamic
  interactions between the droplets in the co-moving frame. The
  transverse forces resulting from leading and trailing droplets in a
  zigzag configuration with one `mismatch' are shown as grey and black
  arrows, respectively. c) Time series of a transverse droplet motion
  for a single `mismatch' as described in b); three drops are marked
  and the net transverse forces, i.e., the sum of the grey and black
  arrows in b), are shown as black arrows.
}
\label{Exp1}
\end{figure}

Another protocol to excite waves in the microfluidic crystal is to
exchange the positions of a pair of droplets. Such `defects' can be
created experimentally by guiding the zigzag arrangement of droplets
around a rectangular microfluidic bend, see figure \ref{Exp1}\,a). At
the bend, the two droplets of a pair exchange their longitudinal
position as depicted. During this process, the passing droplet creates
an accelerated flow at its trailing edge which prevents the following
pair from exchanging positions. Under suitable conditions, every
second pair undergoes a positional reordering such that the
translational symmetry of the droplet train is broken. Hence, in the
resulting droplet arrangement after the bend the droplets experience a
net hydrodynamic force.
%
%
For certain droplet respectively channel dimensions, the bend allows
to systematically create such defects in the translational symmetry of
the crystal, and if defects are created periodically a global
oscillation patterns emerge. 
%
%
The collective oscillations are very stable and could be observed for
channel lengths up to 10\,cm, i.e. after travel distances which are
four orders of magnitude larger than a typical droplet radius. The
wavelength $\lambda$ in longitudinal direction depends on both the
droplet size and the droplet spacing. By a variation of these
parameters, various initial wavelengths can be excited.
\begin{figure}
\centering
    {\includegraphics[width=\columnwidth]{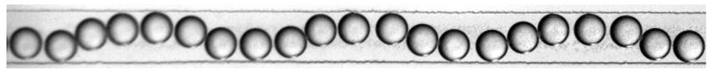}}
    {\includegraphics[width=\columnwidth]{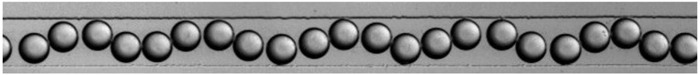}}
    {\includegraphics[width=\columnwidth]{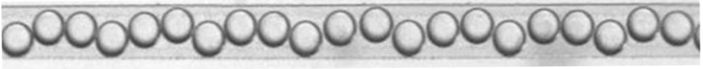}}
\caption{Experimentally observed travelling sine waves as generated by
  periodic `mismatches' using the setup shown in figure \ref{Exp1}
  having different droplet size ($R \approx 64\,\mu\textrm{m}$, $R
  \approx 67\,\mu\textrm{m}$, and $R \approx 70\,\mu\textrm{m}$ from
  top to bottom) and different wavelength ($\lambda=8\,a$,
  $\lambda=6\,a$, and $\lambda=4\,a$ from top to bottom).}
\label{fig:exp_trans}
\end{figure}
\begin{figure}[t]
  {\includegraphics[width=\columnwidth]{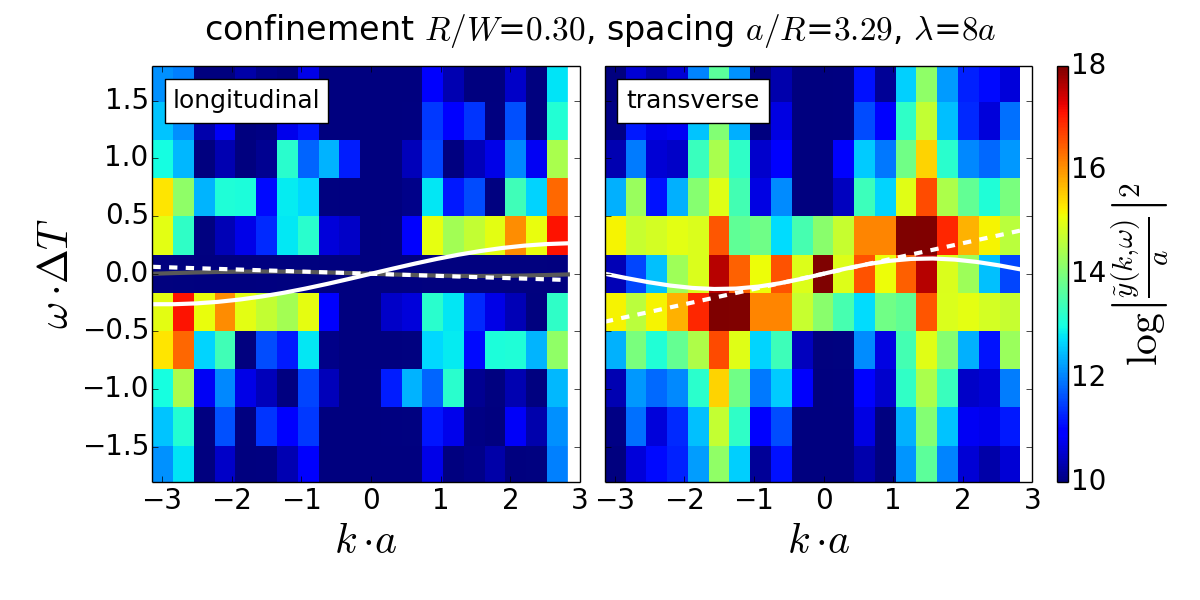}}
  {\includegraphics[width=\columnwidth]{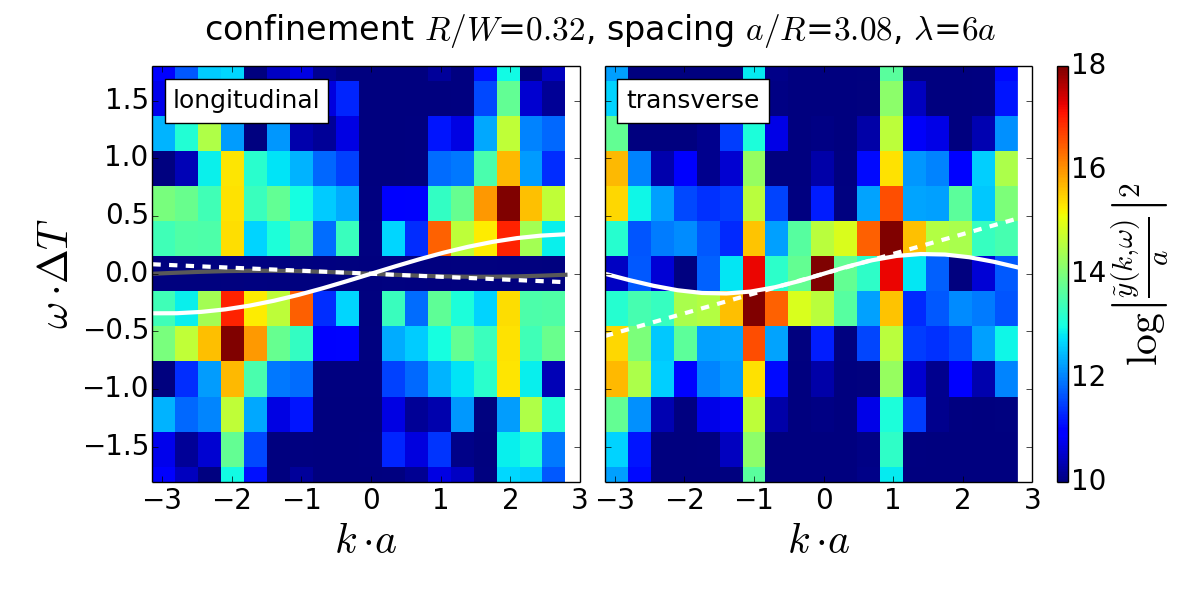}}
  {\includegraphics[width=\columnwidth]{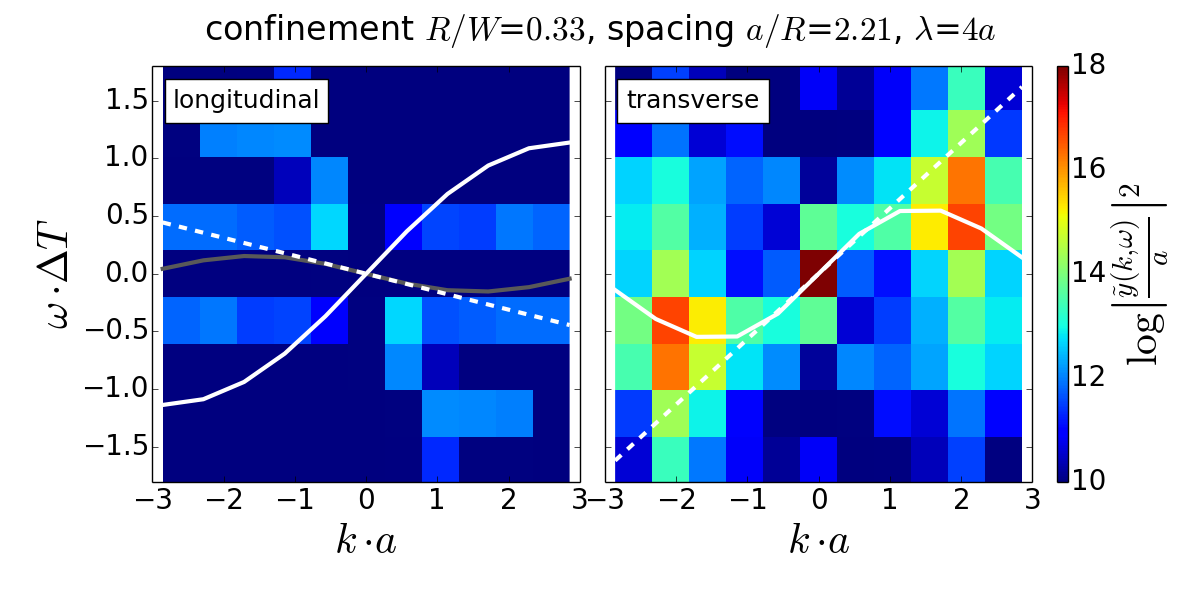}}
\caption{Longitudinal and transverse power spectra of the droplet
  oscillations as calculated from experimental trajectories extracted
  from microscopy time series corresponding to the waves shown in
  Fig.~\ref{fig:exp_trans}.}
\label{Exp2}
\end{figure}
Within a certain parameter range, we can achieve accurate control of
the amplitude of the transverse oscillations by tuning the droplet
radius $R$, since the amplitude of the transverse oscillations is
equal to $W-2R$, where $W$ is the channel
width. Figure~\ref{fig:exp_trans} shows three examples of structures
with variable wavelength and amplitude.  They were obtained from
zigzag structures with equal droplet spacing but with different
droplet radii $R \approx 64\,\mu\textrm{m}$, $R \approx
67\,\mu\textrm{m}$, and $R \approx 70\,\mu\textrm{m}$ which represent
the case of the largest, intermediate, and smallest wavelength
producible. The first structure presents $8$ droplets, the second $6$
droplets, and the last structure $4$ droplets per wavelength. From the
microscopy time series we extract the trajectories and from the
trajectories we could extract the transverse phonon spectra, see
figure \ref{Exp2}, which reveal the presence of two peaks for the
transverse oscillations. These peaks indicate that the excitations
have distinctive wavelengths for both longitudinal and transverse
modes.

\begin{figure}
\includegraphics[width=\columnwidth]{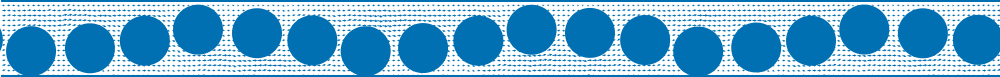}\smallskip\\
\includegraphics[width=\columnwidth]{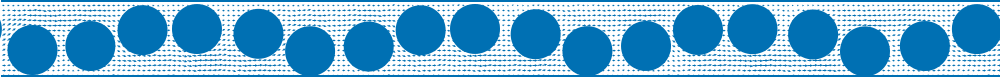}\smallskip\\
\includegraphics[width=\columnwidth]{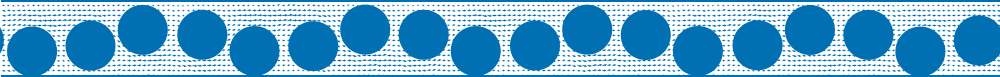}
\caption{Snapshots of sine-like transverse waves observed in computer
  simulations starting from an initial triangle wave.$^\dag$ The
  variable wavelength $\lambda$ in the configurations shown from top
  to bottom are $6\,a$, $5\,a$, and $4\,a$.}
\label{fig:triangle-setup}
\end{figure}
\begin{figure}[ht]
\includegraphics[width=\columnwidth]{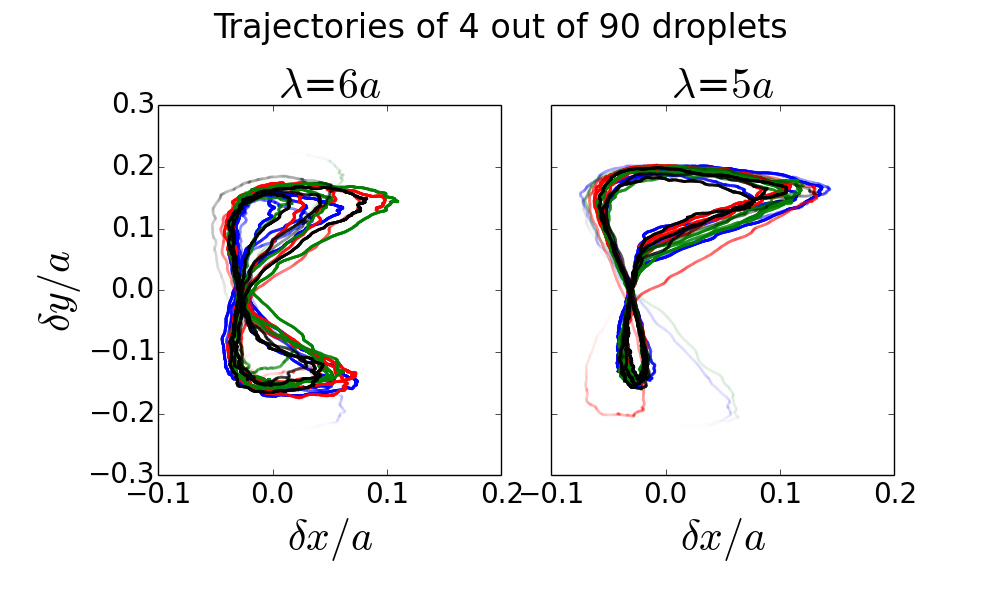}
\caption{Trajectories of $4$ droplets in a train of $90$ droplets
  forming a sine-like wave as obtained from simulations for
  wavelengths $\lambda=6\,a$ and $\lambda=5\,a$. The trajectories
  follow a figure eight-like pattern, which is symmetric for
  $\lambda=6\,a$ and asymmetric for $\lambda=5\,a$.}
\label{fig:trajectories-trans}
\end{figure}

The same type of droplet wave also emerges in computer simulations
when a triangle wave is used as initial condition, see figure
\ref{fig:triangle-setup}. Various wavelengths can be excited which are
found to behave qualitatively the same.
In the following, we analyse results obtained for wavelengths $\lambda
= 5\,a$ and $\lambda = 6\,a$, corresponding to five and six droplets
within one wavelength, respectively. Figure
\ref{fig:trajectories-trans} shows the trajectories of the droplets in
the co-moving frame of reference.  We find that both the longitudinal
and transverse coordinates oscillate around the equilibrium
position. The configuration-space trajectory of each droplet describes
a figure-eight pattern. While for $\lambda = 6\,a$ the figure-eight
pattern is almost symmetric, it is clearly antisymmetric for $\lambda
= 5\,a$. This is a consequence of an antisymmetric initial arrangement
of droplets at the top and bottom walls where two neighbouring
droplets are close to the top walls, while a single droplet is close
to the bottom wall. The trajectories in figure
\ref{fig:trajectories-trans} suggest that this asymmetry is maintained
in the oscillatory motion.$^\dag$
The trajectories are reminiscent of the bound-state motion pattern of
an isolated pair of rigid discs in two-dimensional Poiseuille flow
\cite{Reddig2013}. Quantitatively, however, the state diagrams in
Ref.~\cite{Reddig2013} seem to predict cross-swapping trajectories for
the relatively large displacements ($|y_1-y_2|/W \approx 0.2$ and
$2a/W \approx 1.5$) in the wave configurations considered here. We
conclude that here the ensemble arrangement stabilises the oscillatory
states, as already observed for the pairing waves as discussed in
section \ref{sec:long} for the cascade of droplet pairing.

\begin{figure}[t]
\includegraphics[width=\columnwidth]{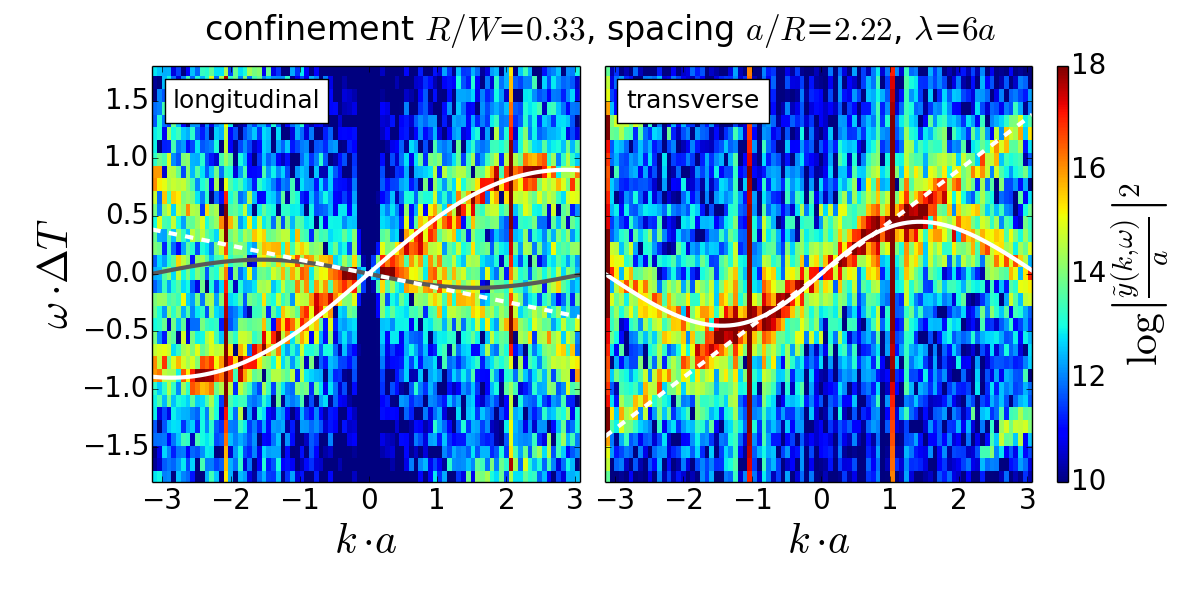}\\
\includegraphics[width=\columnwidth]{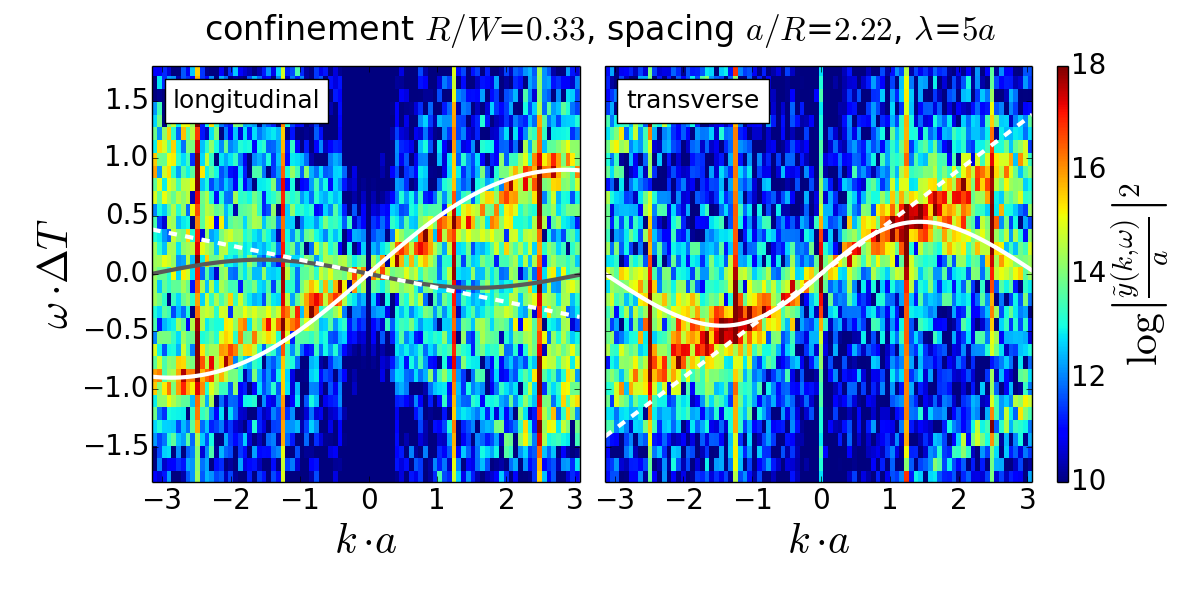}
\caption{Power spectra obtained from simulations of the
  sine-like waves with wavelength $\lambda=6\,a$ (top) and $\lambda=5\,a$
  (bottom). The white lines are the dispersion relations
  $\omega_\perp(k)$ for transverse phonons, Eq.~(\ref{eq:dispersion})
  and the phenomenological dispersion relation $\omega_\parallel(k)$
  from Eq.~(\ref{eq:dispersion3}). The theoretical dispersion relation
  (\ref{eq:dispersion}) for longitudinal phonons is shown in grey. The
  dashed lines are the continuum approximations of
  Eq.~(\ref{eq:dispersion4}).}
\label{fig:spectra-trans}
\end{figure}

\begin{figure}
\includegraphics[width=\columnwidth]{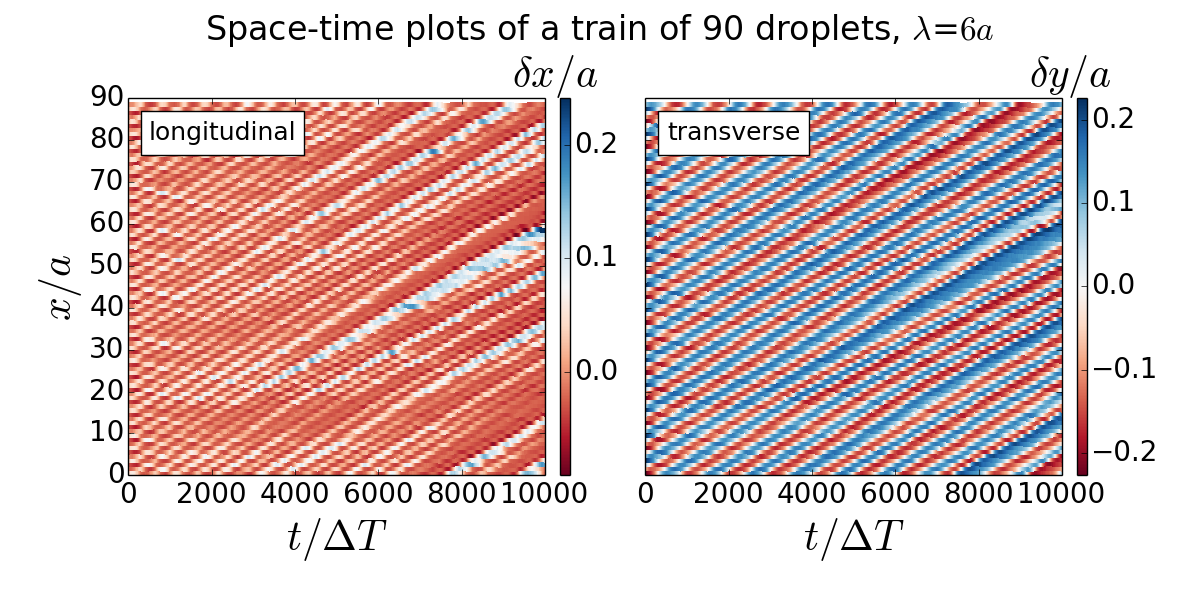}\\
\includegraphics[width=\columnwidth]{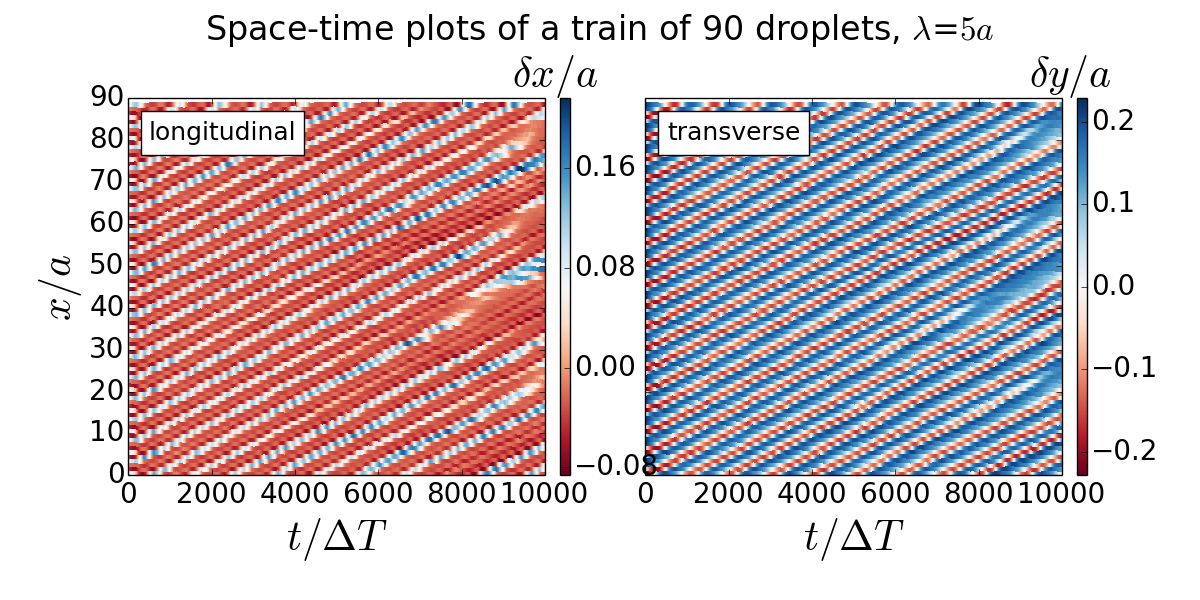}
\caption{Space-time plots of simulation runs of a train of $90$
  droplets forming a sine-like wave for wavelengths $\lambda=6\,a$ and
  $\lambda=5\,a$.
}
\label{fig:spacetime-trans}
\end{figure}

Figure \ref{fig:spectra-trans} shows the longitudinal and transverse
power spectra of the waves depicted in figure
\ref{fig:triangle-setup}. The transverse power spectrum from
simulations shows a continuous signature where the dependence of the
frequency on the wave-vector has a sine-like shape. Such a dispersion
relation is reminiscent of microfluidic
phonons.\cite{Beatus2007,Beatus2012} The dispersion relation predicted
by the linearised far-field theory for confined microfluidic
phonons,\cite{Beatus2009,Beatus2012} cf. section~\ref{sec:theory},
%
%
is plotted on top of the spectra and shows excellent quantitative
agreement. It is worthwhile to note that the calibration procedure
described in section \ref{sec:mpc} fixes all parameters such that no
fitting is needed. 
The space-time plot of the droplet distance in figure
\ref{fig:spacetime-trans} confirms that the wave is travelling in the
flow direction along the droplet crystal, and is considerably more
stable than the longitudinal waves discussed above. At long times,
gaps start forming and the crystal breaks up into smaller sub-units.
These results show that the excited transverse wave is an acoustic
microfluidic phonon, and our experimental approach enables us to
specifically excite such modes with a large amplitude. 
The power spectra also show a clear signature of longitudinal
oscillations which is explained in detail in the following subsection.

\begin{figure}[t]
  {\includegraphics[width=\columnwidth]{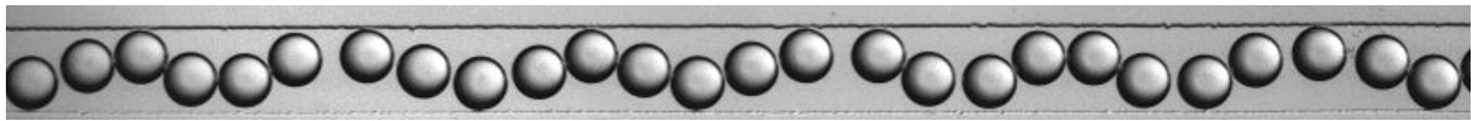}\smallskip}
  \smallskip\\
  {\includegraphics[width=\columnwidth]{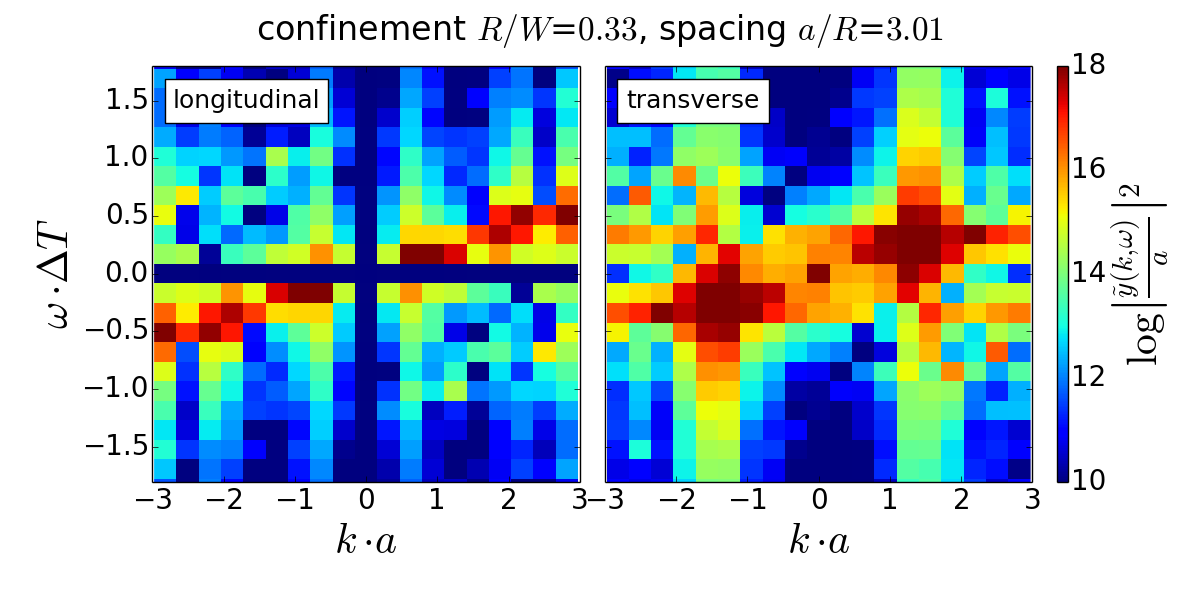}}
\caption{(Top) Experimentally observed droplet pattern as generated by
  periodic `mismatches' using the setup shown in figure \ref{Exp1}
  with increased spacing or `gaps' between some droplets. (Bottom)
  Corresponding longitudinal and transverse power spectra of the
  droplet oscillations as calculated from experimental trajectories
  extracted from microscopy time series.}
\label{Exp3}
\end{figure}

Here it is also possible to increase experimentally the amplitude of
the longitudinal waves by increasing the distance between the
droplets. A droplet train with such `gaps' is shown in figure
\ref{Exp3} along with the corresponding longitudinal and transverse
power spectra.  Compared to the sine-like waves without gaps,
cf. figure \ref{Exp2}, the transverse modes here exhibit a broader
signal around the main wavelength. Moreover, the longitudinal
signature extends over a whole range of wavelength, similar to the
longitudinal modes observed in computer simulations, cf. figure
\ref{fig:spectra-long}. This indicates that a range of wavelengths
have been excited longitudinally by increasing the droplet spacing,
while the transverse sine-like wave is still predominant. From the
optofluidic point of view, these heterogeneous structures are the most
interesting. When the droplet distance approaches the hydrodynamic
screening length, more heterogeneous structures are observed in line
with the results from computer simulations in section \ref{sec:long}.

\subsection{Coupling of longitudinal and transverse oscillations}


The experimental and numerical power spectra in figure \ref{Exp2} and
\ref{fig:spectra-trans} also show a clear signature of longitudinal
oscillations, in particular for the simulation results. The dominant
frequencies from experimental measurements agree quantitatively with
the simulation results. However, the dispersion relation of the
longitudinal modes is qualitatively different from the prediction of
the linearised theory. The frequency curve $\omega_\parallel(k)$ for
acoustic phonons, c.f. Eq. (\ref{eq:dispersion}),
%
%
is plotted as a grey line in figure \ref{fig:spectra-trans} and
describes waves that propagate upstream, whereas the observed
frequencies indicate a positive group velocity. The shape of the
measured frequency curve resembles instead the shape of the
transverse dispersion relation $\omega_\perp$, yet the maximum appears
shifted towards the edge of the Brillouin zone and has a higher
frequency than the maximum of $\omega_\perp(k)$.

The anomalous properties of the longitudinal modes are also apparent
in the correlation strength
\begin{equation}
C(k_\parallel,k_\perp) 
= \frac{ \left\langle \tilde{x}(k_\parallel,t) 
  \tilde{y}(k_\perp,t) \right\rangle_t } 
  { \sqrt{ \left\langle |\tilde{x}(k_\parallel,t)|^2 \right\rangle_t 
  \left\langle |\tilde{y}(k_\perp,t)|^2 \right\rangle_t } }
\end{equation}
between longitudinal and transverse modes

\begin{equation}
\begin{split}
\tilde{x}_\parallel(k,t) &= \sum_{j=1}^{N} \delta x_{j}(t) 
  \exp \left[ -2\pi i \frac{j k}{N} \right] \\
\tilde{y}_\perp(k,t) &= \sum_{j=1}^{N} \delta y_{j}(t)
  \exp \left[ -2\pi i \frac{j k}{N} \right].
\end{split}
\end{equation}

\begin{figure}[t]
\includegraphics[width=\columnwidth]{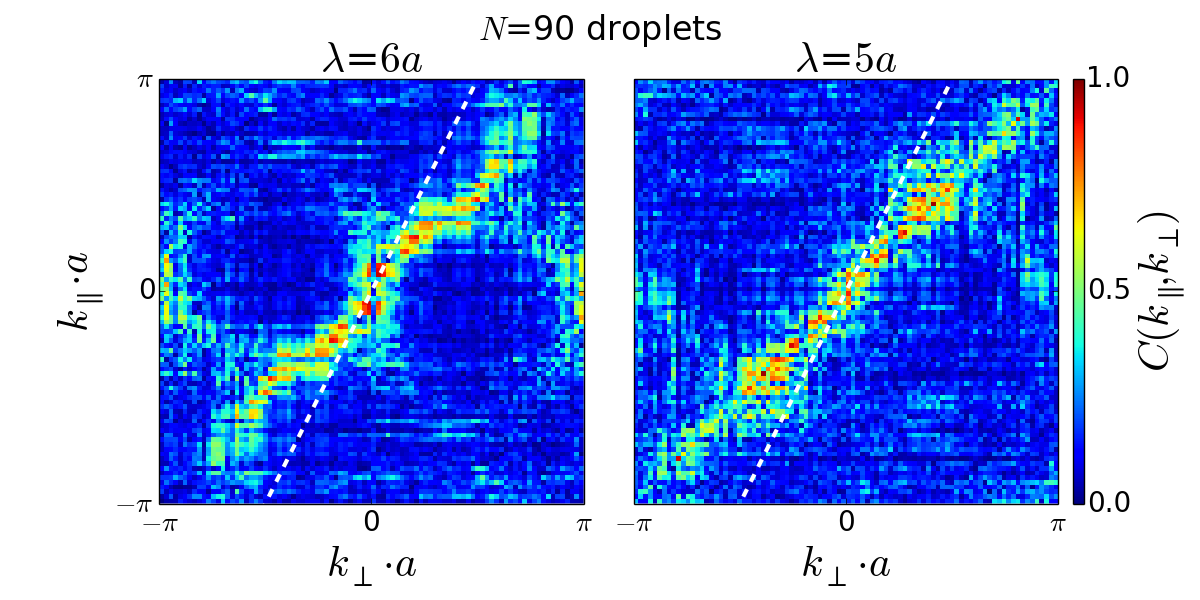}
\caption{Correlation strength $C(k_\parallel,k_\perp)$ of phonon modes
  obtained from simulations of transversely excited oscillations with
  initial wavelengths $\lambda=6\,a$ and $\lambda=5\,a$. The dashed
  line is the phenomenological relation $k_\parallel = 2 k_\perp$.}
\label{fig:correlation-trans}
\end{figure}

The correlation strength for the transversely excited oscillations is
shown in figure \ref{fig:correlation-trans} and indicates that
longitudinal and transverse modes strongly correlate if $k_\parallel$
and $k_\perp$ have the same sign. This is in contrast to correlations
that are expected for purely acoustic phonons, where the matching
condition $k_\parallel = - k_\perp$ is expected.\cite{Liu2012}
Furthermore, by employing the frequencies for correlated wave-vectors
in the observed dispersion relations (figure \ref{fig:spectra-trans}),
we find that the matching waves approximately obey
$\omega_\parallel=2\omega_\perp$. These observations indicate a strong
coupling of the longitudinal modes to the transverse oscillation which
is beyond a linearised far-field theory.
This coupling is induced by the relatively large amplitude of the
transverse waves which brings the droplets close to the wall where the
imposed flow is not uniform but decays due to no-slip boundary
conditions.\cite{Fleury2014} The droplets are slowed down at both
channel walls which leads to the figure-eight trajectories seen in
figure \ref{fig:trajectories-trans}.
In the co-moving frame, the droplets move backwards when they are
close to the wall, and forward when they cross the channel. During one
transverse cycle, two wall approaches take place such that
$$\omega_\parallel=2\omega_\perp$$ in agreement with the matching
condition found above. Inspection of the spatial wave patterns reveals
that a full longitudinal wave extends over each crest or trough of the
transverse wave, hence $$k_\parallel=2k_\perp.$$ Combining these
conditions leads to the dispersion relation
\begin{equation}\label{eq:dispersion3}
\omega_\parallel(k) = 2 \omega_\perp(k/2) .
\end{equation}
This relation is also plotted in figure \ref{fig:spectra-trans} and is
in remarkable quantitative agreement with the signature of the
longitudinal oscillations. 


\begin{figure}[t]
\includegraphics[width=\columnwidth]{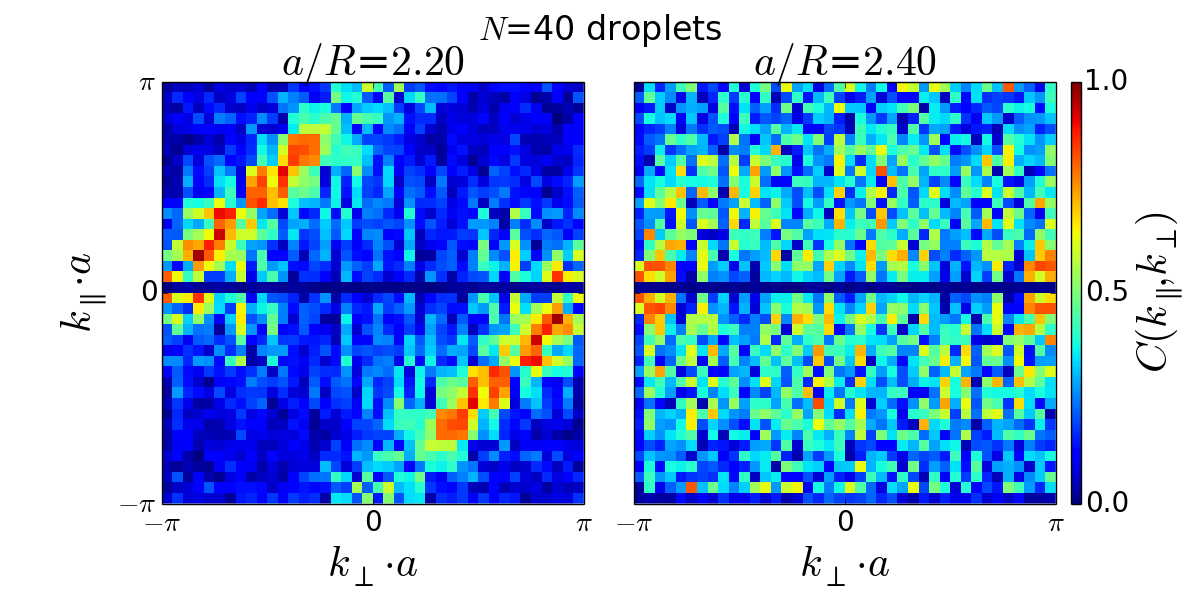}
\caption{Correlation strength $C(k_\parallel,k_\perp)$ obtained from
  simulation data of longitudinal pairing waves triggered by gaps.}
\label{fig:correlation-long}
\end{figure}

The correlation strength for the longitudinal pairing waves discussed
in section \ref{sec:long} is shown in figure
\ref{fig:correlation-long}. Here, the oscillations seem to be highly
correlated in a region where $k_\parallel = k_\perp - \pi$. However,
this pattern was only found if the gaps between pairs were small, and
since the pairing waves are considerably less stable, it is difficult
to clearly identify the origin.

We conclude that the lateral confinement in connection with the
considerable amplitude of the transverse excitations leads to a strong
interaction of longitudinal and transverse waves in confined
microfluidic crystals. It is interesting to note that for the
excitation mechanism studied here, this coupling does not seem to lead
to an instability. A possible explanation is that in the dense droplet
crystal, longitudinal oscillations cannot grow due to steric
constraints between the droplets.


\subsection{Long-time behaviour and stability}
\label{sec:stability}

\begin{figure}[t]
\includegraphics[width=\columnwidth]{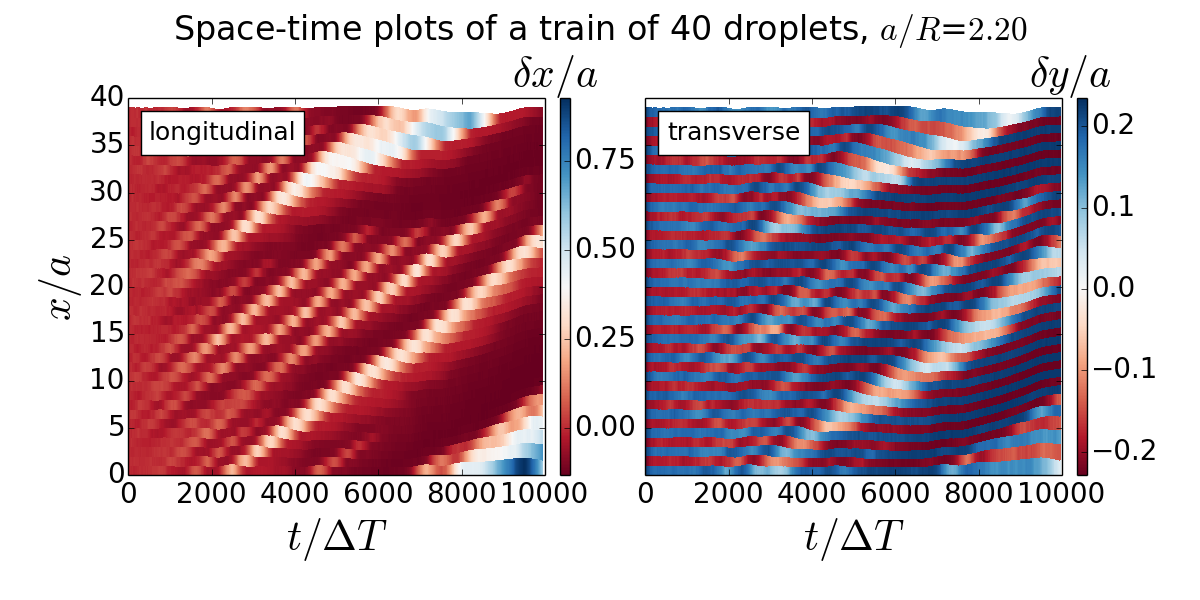}\\
\includegraphics[width=\columnwidth]{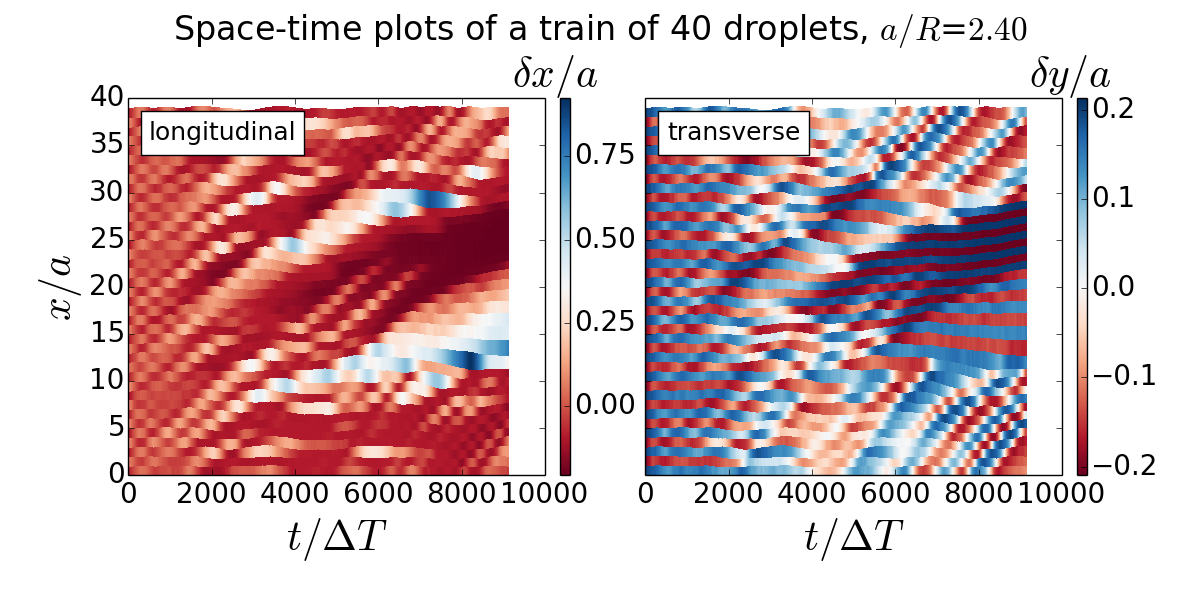}
\caption{Space-time plots of simulation runs of a train of $40$
  droplets initially perturbed by gaps with average spacing between
  the droplets $a/R=2.2$ and $a/R=2.4$, respectively.
}
\label{fig:spacetime}
\end{figure}

\begin{figure*}[t]
\includegraphics[width=\textwidth]{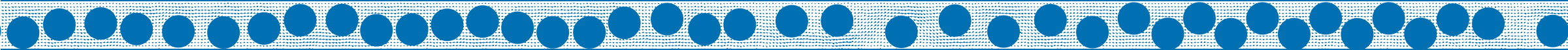}
\caption{Simulation snapshot illustrating the formation of ordered and
  unordered subregions within a train of $40$ droplets. The right part
  of the train has `frozen' into a stable zigzag configuration, while
  the left part exhibits a travelling sine-wave.$^\dag$}
\label{fig:wave-zigzag}
\end{figure*}

While the transverse waves excited by positional exchange appear to be
relatively stable, the pairing waves excited by the introduction of
gaps in the droplet train persist only for much shorter times. This is
evident in the space-time plot of the longitudinal droplet distance in
figure \ref{fig:spacetime}, which extends the time scale of figure
\ref{fig:spacetime-long} by 5 five times and shows that the initial
order disappears and other patterns emerge. One feature that is
visible for a range of initial gap widths is the formation of
subregions with small longitudinal droplet distances (dark red regions
in the space-time plot). Inspection of configuration snapshots reveals
that in these subregions, the droplets arrange in a dense zigzag order
without oscillations. We hypothesise that the formation of these
ordered regions as observed in simulations is similar to flow-induced
crystallisation.\cite{Vermant2005} The `frozen' parts of the droplet
train propagate with a velocity that depends on the size of the
subregion, and due to different velocities some zigzag clusters catch
up and merge with others. Simulation trajectories reveal that the
droplets crystallise at the leading edge and melt at the trailing edge
of the subregions. In between the ordered regions, the droplet train
is disordered. In some instances, we observe a phonon-like transverse
wave that develops in these regions.$^\dag$ One simulation snapshot of
a droplet train that simultaneously exhibits an ordered zigzag region
and a transverse phonon wave is shown in figure \ref{fig:wave-zigzag}.
%

In the simulations, the long-time behaviour is significantly affected
by thermal fluctuations, which tend to lead to a disordered droplet
ensemble where the crystal structure disappears. When the droplets
move away from their regular crystal positions, they can be regarded
as a continuous ensemble, and the finite-differences in the equation
of motion are to be replaced by a continuum approximation
\cite{Rosenau1986, Wattis2000} for small $a$
\begin{equation}
\begin{split}
\begin{split}
  \delta x_{n+j} - \delta x_{n-j}
  & \approx
  2 j a \frac{\partial \delta x}{\partial x}
\end{split}\\
\begin{split}
  \delta y_{n+j}-\delta y_{n-j}
  & \approx
  2 j a \frac{\partial \delta y}{\partial x} .
\end{split}
\end{split}
\end{equation}
Using this continuum approximation to derive the equations of motion
and the dispersion relation as in section \ref{sec:theory}, we obtain
\begin{equation}\label{eq:dispersion4}
\begin{split}
\omega_{\parallel}(k) &= -4 \, k a \, B \, \coth(j
\pi \beta) \mathrm{csch}^2(j \pi \beta) \\
\omega_{\perp}(k) &= 2 \, k a \, B \, \left[ 1 + \cosh(j
\pi \beta) \right]^2 \mathrm{csch}^3(j \pi \beta).
\end{split}
\end{equation}
These relations are plotted as dashed lines on the power spectra in
figure \ref{fig:spectra-trans}. The longitudinal pairing waves clearly
show a dispersion branch that is linear in $k$, and the spectra of the
transverse waves also have a signal that agrees with a linear
dispersion relation. We find that this part of the spectrum becomes
more pronounced at longer simulation times, which supports our
hypothesis that the droplet train becomes disordered due to
fluctuations. Overall our analysis sheds some light on the long-time
behaviour of oscillations in microfluidic crystals, however, more data
will be needed to study the onset and growth of instabilities further.

\section{Summary and discussion}
\label{sec:conclusions}

Excitation mechanisms for collective waves in microfluidic crystals
have been investigated. We have demonstrated that both longitudinal and
transverse waves can be systematically excited by creating specific
defect patterns. Experimental results were confirmed by computer
simulations, and our results reveal instabilities and mode coupling
that originate from the underlying hydrodynamics.

The excited longitudinal modes show cascades of pairs of laterally
displaced droplets. Due to the pairing instability, the pairing
cascade is rather short-lived, and over time other modes are observed.
Since the constraints prevent the dense crystal from becoming
completely disordered, some parts `freeze' into a zigzag arrangement
while others exhibit transverse oscillation. This is a possible
indication of flow-induced crystallisation, and it will be interesting
to further investigate the dynamic formation of the inhomogeneous
regions.

The transverse waves show the dispersion relation of a microfluidic
phonon. Comparison with the analytical prediction demonstrates that a
linearised far-field theory works well even in a dense droplet
crystal. A possible reason is the relatively strong lateral
confinement, which may screen the higher-order reflections of the
dipolar interactions. The power spectra of the oscillations exhibit a
correlation that arises from a coupling of longitudinal to transverse
modes. This coupling is induced by the boundary conditions at the
confining channel walls. At large amplitudes, the inhomogeneity of the
imposed flow affects the dynamics of the droplets and leads to novel
collective modes. Our results thus shed light on the mechanisms underlying
non-linear mode coupling in microfluidic crystals.

Experimentally, the excited waves are highly stable and do not undergo
instabilities. This raises the question whether some of the
instabilities observed in microfluidic crystals
\cite{Beatus2012,Janssen2012} are suppressed in the dense droplet
ensembles studied here. Large displacements are inhibited by steric
interactions between the droplets which may have a stabilising effect,
e.g., on the pairing cascades. A detailed study of the impact of
geometric constraints on the stability of the collective modes is a
topic for future research. In the simulations, the smaller Peclet
number leads to a more pronounced influence of thermal
fluctuations. This opens up the possibility for fluctuation-induced
instabilities and our simulation approach may thus be used to
investigate instabilities in microfluidic devices.

The good agreement of the experimental and numerical results with the
linearised far-field prediction suggests that the dynamics of dense
and confined microfluidic droplets is accessible theoretically and
leads to novel insights into the origin of instabilities and mode
coupling effects. Furthermore, it seems promising to apply the
experimental and numerical techniques to other microfluidic systems,
such as dense droplet systems in 2D, crowded particle systems, or
self-propelled particles. Finally, the experimental techniques may be
used to control particle flows in microfluidic applications such as
flow cytometry or high-throughput assays using microchips.



\section*{Acknowledgements}

The authors would like to thank Tsevi Beatus, Itamar Shani, Roy
Bar-Ziv and Roland G. Winkler for valuable discussions. R.S. and
J.-B.F. gratefully acknowledge financial support by the DFG grant SE
1118/4.





\footnotesize{
\bibliography{droplets} 
\bibliographystyle{rsc} 
}


\end{document}